\documentclass[12pt,preprint]{aastex}
\usepackage[margin=0.8in]{geometry}
\usepackage{epsfig,xspace}
\usepackage{graphicx, wrapfig}
\usepackage{wrapfig}
\usepackage{units}
\usepackage{amsmath}
\usepackage{amsfonts}
\usepackage{url}

\usepackage{natbib}
\newcommand{\captionfonts}{\normalsize}

\makeatletter  
\long\def\@makecaption#1#2{%
  \vskip\abovecaptionskip
  \sbox\@tempboxa{{\captionfonts #1 #2}}%
  \ifdim \wd\@tempboxa >\hsize
    {\captionfonts #1 #2\par}
  \else
    \hbox to\hsize{\hfil\box\@tempboxa\hfil}%
  \fi
  \vskip\belowcaptionskip
  }
\makeatother   

\newcommand{\bt}[1]{\textbf{#1}}

\newcommand{\m}{\text{m}}

\newcommand{\msun}{\ensuremath{{M}_\odot}}

\newcommand{\swift}{\textit{Swift}}

\begin{document}

\title{The Evolution of the Far-UV Luminosity Function and Star Formation Rate Density of the Chandra Deep Field South from \bt{\textit{z}}=0.2-1.2 with Swift/UVOT}

\author{
Lea M.~Z.\ Hagen\altaffilmark{1,2}, 
Erik A.\ Hoversten\altaffilmark{3}, 
Caryl Gronwall\altaffilmark{1,2}, 
Christopher Wolf\altaffilmark{1}, 
Michael H.\ Siegel\altaffilmark{1}, 
Mathew Page\altaffilmark{4},
and
Alex Hagen\altaffilmark{1,2}
}

\affil{\altaffilmark{1}Department of Astronomy and Astrophysics, The 
Pennsylvania State University, University Park, PA 16802, USA} 
\affil{\altaffilmark{2}Institute for Gravitation and the Cosmos, The 
Pennsylvania State University, University Park, PA 16802, USA} 
\affil{\altaffilmark{3}Department of Physics \& Astronomy, University of North Carolina-Chapel Hill, 120 E.\ Cameron Ave., Chapel Hill, NC 27599, USA}
\affil{\altaffilmark{4}Mullard Space Science Laboratory, University College London, Holmbury St Mary, Dorking, Surrey RH5 6NT, UK}

\email{lmz5057@psu.edu}



\begin{abstract}

We use deep \swift\ UV/Optical Telescope (UVOT) near-ultraviolet (1600~\AA\ to 4000~\AA) imaging of the Chandra Deep Field South to measure the rest-frame far-UV (FUV; 1500 \AA) luminosity function (LF) in four redshift bins between $z=0.2$ and 1.2.  Our sample includes 730 galaxies with $u < 24.1$ mag.
We use two methods to construct and fit the LFs: the traditional $V_\text{max}$ method with bootstrap errors and a maximum likelihood estimator.   We observe luminosity evolution such that $M^*$ fades by $\sim 2$ magnitudes from $z \sim 1$ to $z \sim 0.3$ implying that star formation activity was substantially higher at $z\sim 1$ than today.
We integrate our LFs to determine the FUV luminosity densities and star formation rate densities from $z=0.2$ to $1.2$.  We find evolution consistent with an increase proportional to $(1+z)^{1.9}$ out to $z \sim 1$.  Our luminosity densities and star formation rates are consistent with those found in the literature, but are, on average, a factor of $\sim$2 higher than previous FUV measurements. In addition, we combine our UVOT data with the MUSYC survey to model the galaxies' ultraviolet-to-infrared spectral energy distributions and estimate the rest-frame FUV attenuation.  We find that accounting for the attenuation increases the star formation rate densities by $\sim$1~dex across all four redshift bins.

\end{abstract}


\section{Introduction}

Establishing the evolution over cosmic time of the star formation rate density of the universe provides crucial constraints for current models of galaxy formation and evolution \citep[e.g.,][]{somerville12}.  
Previous work has shown that the volume-averaged star formation rate density (SFRD) has increased between now and $z \approx 1$,  flattened between $z=1$ and 4, and decreased for $z>4$ \citep[e.g.,][]{lilly96,hopkins06}.  The details of this evolution, however, are not well understood, due to
(a) the variety of star formation rate (SFR) indicators used, which have associated systematic uncertainties;
(b) uncertainties arising from cosmic variance due to the relatively small volumes probed by any individual observational estimate of the SFRD in a given redshift bin;
and (c) complex selection criteria that can be difficult to account for in the calculated SFRD uncertainties.

While there are a variety of SFR estimators used in the literature \citep[see][for a review]{kennicutt12},
the ultraviolet (UV) light is one of the most direct as the UV light
emitted by young massive stars dominates the spectral energy distributions of newly-formed stellar populations.  Far-UV light ($\sim$1500~\AA) is present for $\sim$100~Myr, and thus provides a particularly useful probe of recent star formation. 
The disadvantage of using UV as a SFR tracer is that it is strongly extinguished by dust and the dust extinction law in the ultraviolet is not well understood.
There are many surveys that have probed the UV light emitted by galaxies in the nearby universe ($z\lesssim1.5$)
\citep[e.g.,][]{treyer98,sullivan00,gabasch04,wyder05,schiminovich05,tresse07,oesch10,robotham11,cucciati12}.
At these lower redshifts, one can either probe the rest-frame near-UV emission using optical telescopes, or probe rest-frame far-UV using observations in the near-UV.

Because of the limited availability of wide-field
ultraviolet  telescopes, only a handful of fields have been observed in rest-frame far-UV to sufficient depth to measure the faintest galaxies \citep{wyder05, schiminovich05, arnouts05, robotham11}.  Therefore, calculations of luminosity functions and star formation rate densities are subject to cosmic variance issues.  Additional fields will help to reduce the importance of cosmic variance as a source of uncertainty \citep{madau14}.  Also, measurements utilizing \textit{GALEX} \citep[Galaxy Evolution Explorer;][]{martin05} observations are susceptible to confusion, and improvements upon its $5''$ resolution will lead to cleaner estimates of the SFR density.

We address these needs using deep observations from the UV/Optical Telescope \citetext{UVOT; \citealt{roming05}} on \swift\ \citep{gehrels04}
of the Chandra Deep Field South \citetext{CDF-S; \citealt{giacconi02}}.  The UVOT observations cover observed-frame wavelengths of 1600-4000~\AA\ with a total exposure time of 500~ks, at a resolution of $2.5''$.
Using these data, we construct rest-frame FUV luminosity functions in four redshift bins between $z=0.2$ and 1.2, and use these to calculate the respective star formation rate densities.  This is the first time that UVOT data have been used to construct a history of star formation in the universe.
We also combine the UVOT data with optical and infrared (IR) observations from MUSYC \citep{cardamone10} and model the UV-to-IR spectral energy distributions to find accurate FUV dust attenuations.  The multi-filter NUV coverage of UVOT provides stronger constraints on the rest-frame UV spectral slope -- and thus the FUV attenuation \citep[e.g.,][]{meurer99} -- than does the single \textit{GALEX} NUV filter.

 In \S\ref{sec:data} we describe our sample of galaxies, which are corrected for various biases in \S\ref{sec:corr}.  
We model the spectral energy distributions in \S\ref{sec:sed_fitting}, using the models to determine the FUV dust attenuation.

In \S\ref{sec:lf} we derive the luminosity functions and fit them with Schechter functions \citep{schechter76}, and then calculate SFR densities in \S\ref{sec:sfr}.  We conclude in \S\ref{sec:con}.  Throughout this paper, we use flat $\Lambda$CDM cosmology with $\Omega_M = 0.27$, $\Omega_\Lambda = 0.73$, and $h = 0.71$.  Magnitudes are given in the AB system \citep{oke74}.


\section{Data} \label{sec:data}

Observations of the CDF-S were made with UVOT \citep{roming05}, one of three telescopes on board the \swift\ spacecraft \citep{gehrels04}.  UVOT is a 30 cm telescope with two grisms and seven broadband filters, four of which are used here.  These four near-UV filters and their properties are listed in Table~\ref{filter_data}.
For a detailed discussion of the filters, as well as plots of the responses, see \citet{poole08} and updates in \citet{breeveld11}.
The observations were made between 2007 July~7 and 2007 December~29.  All observations were taken in unbinned mode, with a pixel scale of $0.5''$.  Total exposure times and image areas in each filter are also in Table~\ref{filter_data}. 

\begin{deluxetable}{lcccccc}
\tablecaption{\swift\ UVOT Observations of the CDF-S \label{filter_data}}
\tablewidth{0pt}
\tabletypesize{\footnotesize}
\tablehead{
   \colhead{Filter} & \colhead{Central Wavelength} & \colhead{FWHM} 
   & \colhead{PSF FWHM} & \colhead{Exposure} 
   & \colhead{Area} \\
   \colhead{} & \colhead{(\AA)} & \colhead{(\AA)} 
   & \colhead{} & \colhead{(s)} 
   & \colhead{(arcmin$^2$)}
}
\startdata
uvw2 & 1928 & 657 & $2.92''$ & 144763 & 271.3 \\
uvm2 & 2246 & 498 & $2.45''$ & 136286 & 268.4 \\
uvw1 & 2600 & 693 & $2.37''$ & 158334 & 269.1 \\
u        & 3465 & 785 & $2.37''$ & 124787 & 266.0 \\
\enddata

\tablecomments{UVOT filters and exposures in the CDF-S.  The filters' central wavelengths (the midpoint between the half-maximum wavelengths), FWHMs, and image PSFs are from \citet{breeveld10}. Image area was determined by where the exposure time was at least 50\% of the maximum exposure time. }

\end{deluxetable}

The UVOT data reduction followed that described in \citet{hoversten09,hoversten11}; UVOT data processing
is described in the UVOT Software Guide.\footnote{http://heasarc.gsfc.nasa.gov/docs/swift/analysis}
Exposure maps and images were generated with UVOT FTOOLS (HEAsoft 6.6.1).\footnote{http://heasarc.gsfc.nasa.gov/docs/software/lheasoft/}
This involves two flux conserving interpolations of the images;
the first of these converts from the raw frame to sky coordinates, and the second occurs when summing the images.  During processing, a correction is applied for known bad pixels.

The UVOT detector is a photon-counting device, so as a result, it is subject to coincidence loss.  If more than one photon lands in approximately the same location within the 11 ms readout time, it will only be counted as one detection \citep{fordham00}.  Coincidence loss is only important at the 1\% level for $m_\text{AB} \sim 19$; our objects are sufficiently faint that this effect is insignificant, and no corrections are made.
 
Cosmic ray corrections are not necessary for UVOT images.
Individual events are identified and centroided upon in each
UVOT frame and placed into an image at a later stage. A 
cosmic ray hitting the detector will register one or a few counts after
centroiding, rather than the thousands of counts which occur in
CCDs operating in the usual integrating modes. As a result, 
cosmic rays are part of the background in UVOT images.

Galaxies were identified in the UVOT image using Source Extractor \citep[SE; version 2.5.0;][]{bertin96} and processed in a manner identical to that described in \citet{hoversten09}.
SE generated the background map, which estimates the local background due to the sky and other sources.  The filtering option was used to improve the detection of faint extended sources; the chosen Gaussian filter had a full width at half-maximum (FWHM) identical to that of the PSF of each image.  Galaxy magnitudes were calculated from \verb=MAG_AUTO=, which is designed to give the best total magnitudes for galaxies, and converted to AB magnitudes.

Our galaxy sample was selected based on detections in the $u$ filter.  We only include objects where the exposure time was at least half the maximum exposure time; \swift\ observes with different roll angles, so the field orientation changes with each image, leading to a non-uniform exposure time.
Redshifts for each UVOT object were determined using MUSYC \citep{cardamone10} survey data from Subaru and \textit{Spitzer} IRAC imaging.  MUSYC includes data for the CDF-S in 32 medium and wide photometric bands, spanning a wavelength range of 3500~\AA\ to 8~$\mu\m$.  The resulting spectral energy distributions allow reasonable calculations of galaxies' photometric redshifts.  Over our redshift range, the redshifts are typically good to $\sigma_z/(1+z) \approx 0.007$, with a catastrophic failure rate of $\sim$4\%.

\begin{figure}[h!]
    \centering
       \includegraphics[trim = 10mm 20mm 10mm 20mm, clip=true, scale=0.6, angle=180]{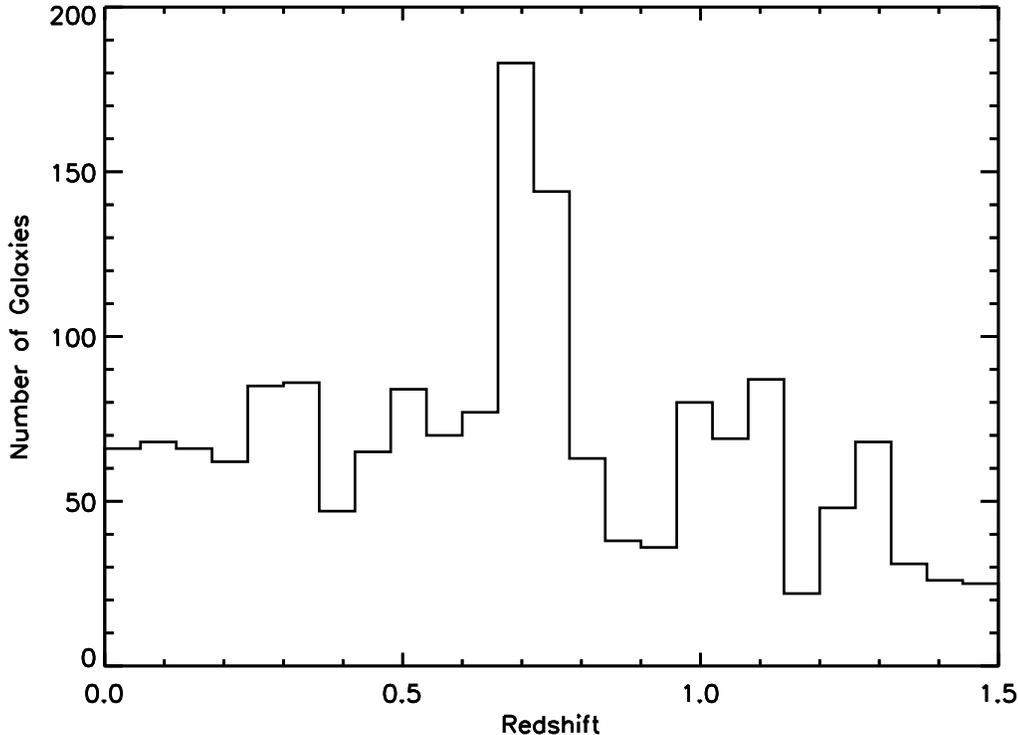}
    \caption{The distribution of redshifts in our CDF-S galaxy sample, as found by matching to MUSYC data.  The peak at $z\approx 0.7$ is a previously known overdensity in this field \citep{gilli03}.}
    \label{z_dist}
\end{figure}

To match objects, UVOT positions were compared with objects in the MUSYC catalog.  If there were multiple objects within $2''$ of a UVOT-detected galaxy, the UVOT and MUSYC spectral energy distributions (SEDs) were compared, and the MUSYC SED with the smallest discontinuity between it and the UVOT SED was chosen as the match.
The resulting distribution of redshifts is in Figure~\ref{z_dist}.  The peak at $z \approx 0.7$ is due to two known galaxy clusters at $z=0.67$ and $z=0.73$ \citep{gilli03}.

\begin{figure}[h!]
    \centering
    \includegraphics[trim = 0mm 25mm 0mm 20mm, clip=true, scale=0.6, angle=180, page=4]{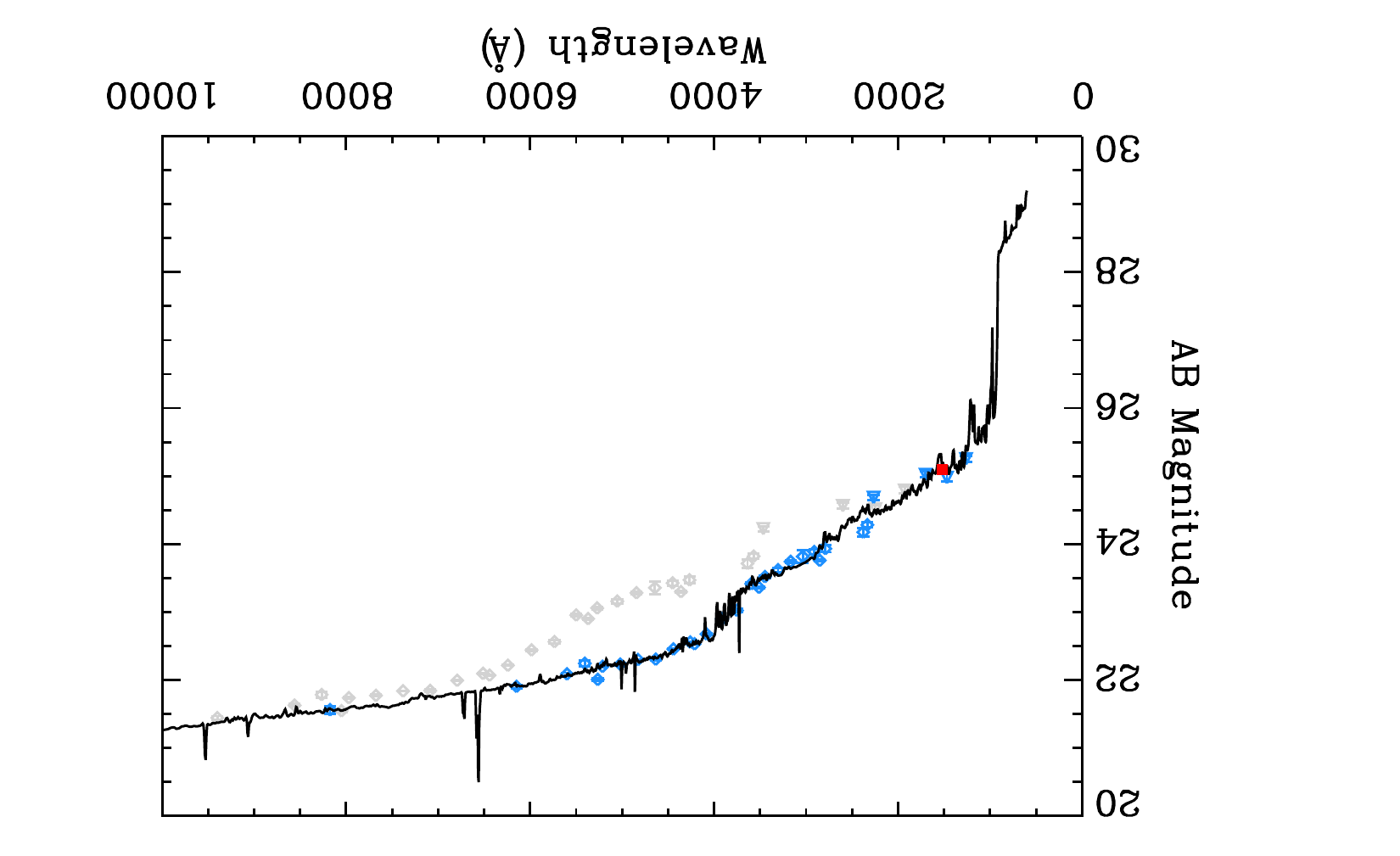}
    \caption{Example of how we derive rest-frame FUV magnitudes for each galaxy.  Grey points are the observed-frame data for a galaxy at $z\approx 0.5$.  Blue points represent the galaxy's rest-frame data, which is then fit with a spectrum.  The red square is the rest-frame \textit{GALEX} FUV magnitude extracted from the spectrum.  It is important to note that the UVOT data extend into the rest-frame FUV, which cannot be done with optical observations for these redshifts.}
    \label{kcorrect}
\end{figure}

To facilitate comparisons to previous work, we determine the rest-frame FUV flux for each galaxy in the field.  To this end, we use \verb=kcorrect= \citep[version 4.2;][]{blanton07}, a software package that fits template spectra to photometric data using nonnegative matrix factorization.  We use the UVOT and MUSYC photometric data to represent the galaxies' spectral energy distribution.  After the software fits a spectrum to each galaxy, it extracts the rest-frame FUV magnitude.  An example of this process for a $z\approx 0.5$ galaxy is in Figure~\ref{kcorrect}.

\section{Bias Corrections} \label{sec:corr}

The data suffer from several biases, which must be corrected before the data are analyzed.  
The first is completeness, in which an object may not be detected due to confusion or photometry limitations.  Due to UVOT's moderate angular resolution, confusion is a small source of error.  It is worth noting that for \textit{GALEX} UV images, which have $5''$ resolution, the incompleteness due to confusion is 21\% \citep{ly09}. 
Second, there is Eddington bias \citep{eddington13}, in which the magnitude errors will preferentially scatter objects into brighter magnitude bins.

These two biases are quantified using a Monte Carlo simulation, following the procedures of \citet{smail95} and \citet{hoversten09}.  Synthetic galaxies with exponential profiles were randomly placed on the UVOT image and the photometry process was repeated.  The distributions of synthetic galaxy magnitudes, semi-major axes, and ellipticities followed those of the original SExtractor results.  The individual photon detections for each galaxy were modeled using Poisson statistics. The profile was convolved with the relevant UVOT PSF before being added into the image.

\begin{figure}[h!]
    \centering
    \includegraphics[trim = 10mm 80mm 30mm 30mm, clip=true, scale=0.6]{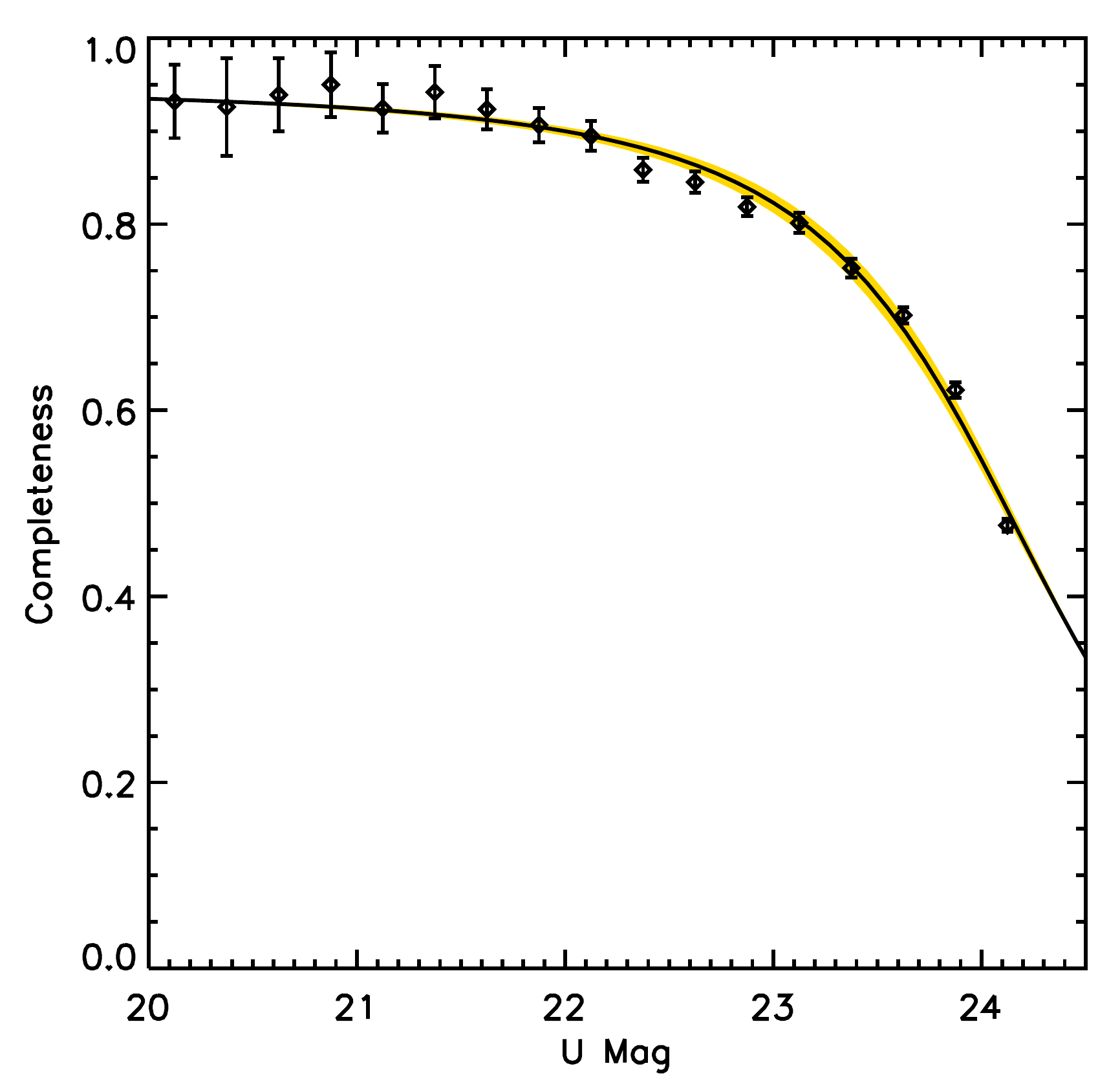}
    \caption{The completeness of detected galaxies as a function of measured $u$ magnitude, as derived with Monte Carlo simulations.  The curve is the best-fit Fleming function \citep{fleming95}, with the fit uncertainty in yellow.}
    \label{completeness}
\end{figure}

In each case, a single synthetic galaxy was randomly added to the original image and the photometry was repeated.  The resulting catalog was checked to determine if the synthetic galaxy was found, and if so, at what magnitude.  The process was repeated approximately 40,000 times.  This yielded an estimate of the completeness as a function of observed magnitude, shown in Figure~\ref{completeness}.  Fainter galaxies were preferentially added to improve the statistics at faint magnitudes.

To make the completeness data more smooth, we fit it with a function of the form used by \citet{fleming95}.  Due to confusion limits, our maximum completeness is about 95\%, so we adjusted the equation accordingly, to
\begin{equation}
C = 0.95 \times 0.5 \left( 
1 - \frac{\alpha(M - M_{50})}{\sqrt{1 + \alpha^2 (M - M_{50})^2}}
\right) ,
\end{equation}
where $M$ is the observed $u$ magnitude, $M_{50}$ is the magnitude corresponding to half the maximum completeness, and $\alpha$ is the steepness of the completeness curve in the vicinity of $M_{50}$.
We fit for the latter two parameters, and find $M_{50} = 24.17 \pm 0.02$ and $\alpha = 0.92 \pm 0.04$.  The best-fit curve is included in Figure~\ref{completeness}.
From this procedure, our sample is 93\% complete to $u = 20$, 80\% to $u=23.1$, and 50\% to $u=24.1$. 

We only considered objects brighter than the 50\% completeness limit.  With this constraint, our $u$-selected sample consists of 1017 galaxies, of which 730 are between redshifts of 0.2 and~1.2.


\section{Spectral Energy Distribution Fitting} \label{sec:sed_fitting}

We combine the UVOT and MUSYC data for our selected galaxies and fit their spectral energy distributions (SEDs).  We use GalMC \citep{acquaviva11}, which utilizes a Markov-Chain Monte Carlo approach.  It fits SEDs over a range of 0.15 to 3~$\mu$m.  We use the Charlot and Bruzual 2007 stellar population synthesis models \citep{bruzual03} and assume a \citet{salpeter55}  initial mass function with M$_\text{min} = 0.1$~\msun\ and M$_\text{max} = 100$~\msun.  We use the \citet{calzetti00} reddening law and account for absorption by the intergalactic medium using \citet{madau95}.  The metallicity is fixed at solar.  Five percent photometric errors were added in quadrature to the known errors in order to account for the error in absolute calibration.  We assume a constant star formation history and fit for three free parameters: stellar mass, the time since the onset of star formation, and E(B-V).

Calculating the galaxies' internal dust extinctions is challenging, due to the lack of certainty in dust extinction laws (discussed further in \S\ref{sec:con}).  However, it is an important part of knowing the true UV luminosities of the galaxies in our sample.  We calculate the expected FUV attenuation ($A_\text{FUV}$) from the \citet{calzetti00} obscuration relation using the modeled E(B-V).

\begin{figure}[h!]
    \centering
    \includegraphics[trim = 15mm 80mm 15mm 60mm, clip=true, scale=0.6]{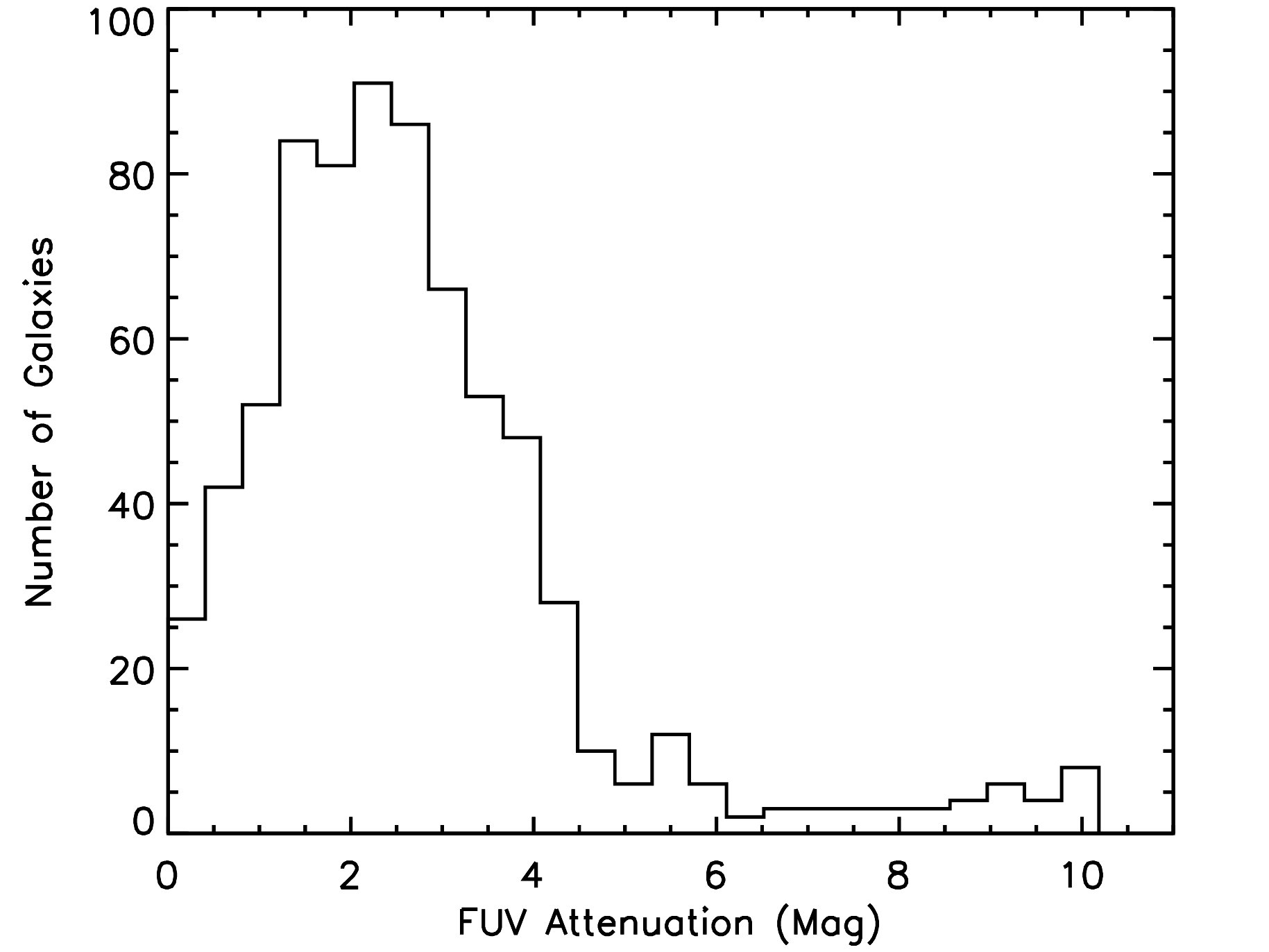}
    \caption{Histogram of the dust attenuation corrections as calculated from SED fitting.}
    \label{fig:ext}
\end{figure}

A histogram of the resulting attenuations is in Figure~\ref{fig:ext}.  We find that 55\% of the galaxies fall within $1 \leq A_\text{FUV} \leq 3$, with a long tail extending to $A_\text{FUV} \approx 10$.  The former galaxies have typical extinction uncertainties that are much smaller than those of the latter galaxies ($\delta A_\text{FUV} \sim$ 0.05~mag versus $\sim$1~mag), so it is not clear that the high extinction values are reliable. 

In the literature, it is common to calculate the average attenuation for redshift bins, and apply that to the value for $M^*$ found in the fits to the uncorrected data.  Following this example, the average FUV attenuation values are in Table~\ref{dust_corr_table}.  However, it is known that attenuation is larger for galaxies with higher SFRs \citep[i.e.,][]{hopkins01, ly12, momcheva13, dominguez13, ciardullo13}.  In addition, because our galaxies are UV-selected, we are missing the most extinguished galaxies.  All other work with UV or optical selection criteria suffers from the same bias.  It is not clear how to correct for this, since both the amount of dust and the proper extinction law are uncertain.  Therefore, for the remainder of this paper, in order to directly compare to results in the literature, we only use data that have not been corrected for dust, unless otherwise specified.

\begin{deluxetable}{cc}
\tablecaption{Average FUV Attenuation \label{dust_corr_table}}
\tablewidth{0pt}
\tabletypesize{\footnotesize}
\tablehead{
\colhead{Redshift} & \colhead{FUV Attenuation} \\
\colhead{} & \colhead{(AB mag)}
}

\startdata
$0.2-0.4$ & 2.26 $^{+1.45} _{-1.45} $ \\
$0.4-0.6$ & 2.28 $^{+1.30} _{-1.30} $ \\
$0.6-0.8$ & 2.29 $^{+1.67} _{-1.24} $ \\
$0.8-1.2$ & 2.35 $^{+1.50} _{-1.25} $ \\
\enddata

\end{deluxetable}



\section{Luminosity Function} \label{sec:lf}

We measure the luminosity function (LF) in two different ways.  For the first of these, we use the traditional $V_\text{max}$ method \citep{schmidt68} to derive the binned data points, with uncertainties determined from a bootstrap technique.  We fit a Schechter function \citep{schechter76} to these data with a chi-squared fitting routine.  Our second method utilizes maximum likelihood estimation (MLE) to find the best-fit Schechter function parameters.


\subsection{$V_\text{max}$ Method} \label{sec:vmax}

The $V_\text{max}$ method for calculating the LF is mathematically expressed as
\begin{equation}
\phi(M) dM
=
\sum_i \frac{1}{C_i V_{\text{max,}i}} , 
\end{equation}
where $\phi(M) dM$ is the number of galaxies with an absolute magnitude between $M$ and $M+dM$ per Mpc$^3$, $C_i$ is the completeness for a galaxy's apparent magnitude (found using the best-fit Fleming function), and $V_\text{max}$ is the maximum volume in which the galaxy could be observed.  

To calculate $V_\text{max}$ for a given galaxy, we first find the range in its observable distance.  The minimum distance is the distance such that our bright end cutoff would have an absolute magnitude equal to that of the galaxy.  The maximum distance is defined identically, but using the faint end cutoff.
The distance range is further constrained to be within the given redshift bin range.  We then calculate the volume of the spherical shell bounded by these distances and the angular area of the image.

The initial error estimate for each data point was calculated using a bootstrap method \citep{efron79}.  With this method, one draws a sample of $N$ objects from a data set of size $N$, with replacement (meaning there will be duplicates of some objects).  One then calculates the quantity of interest.  After repeating the procedure many times, the uncertainties of the quantity are derived from the resulting distribution.  In our case, we randomly chose 730 galaxies (the total number in our sample), with replacement, from the data set.  From this set of galaxies, we calculated $\phi$.  We then repeated the process 500 times.  For each magnitude bin, the median $\phi$ was chosen, with an error defined by the RMS scatter about the median.  This procedure yields more realistic errors for the $\phi$ values than the formal $V_\text{max}$ error, 
\begin{equation}
\sigma \left[ \phi(M) dM \right]
=
\sum_i \frac{1}{C_i^2 V_{\text{max,}i}^2} .
\end{equation}
When there are a small number of galaxies in a magnitude bin, this formulation underestimates the error, which is most pronounced when there is only an upper limit.  The bootstrap method accounts for these situations appropriately. The resulting data points are tabulated in Table~\ref{lf_data}. 

\begin{deluxetable}{c cccc}
\tablecaption{Luminosity Function Data \label{lf_data}}
\tablewidth{0pt}
\tabletypesize{\footnotesize}
\tablehead{
\colhead{FUV Magnitude} & \colhead{$z$=0.2-0.4} & \colhead{$z$=0.4-0.6} & \colhead{$z$=0.6-0.8} & \colhead{$z$=0.8-1.2}
}
\startdata
-22.5 to -22.0 & \nodata & \nodata & \nodata & -4.57 $ ^{+0.20} _{-0.39} $ \\
-22.0 to -21.5 & \nodata & \nodata & \nodata & -4.33 $ ^{+0.16} _{-0.25} $ \\
-21.5 to -21.0 & \nodata & -4.37 $ ^{+0.30} _{-1.82} $ & -4.28 $ ^{+0.24} _{-0.58} $ & -3.92 $ ^{+0.11} _{-0.14} $ \\
-21.0 to -20.5 & \nodata & \nodata & -3.57 $ ^{+0.12} _{-0.17} $ & -3.37 $ ^{+0.07} _{-0.08} $ \\
-20.5 to -20.0 & \nodata & -3.60 $ ^{+0.15} _{-0.23} $ & -2.96 $ ^{+0.07} _{-0.08} $ & -2.98 $ ^{+0.05} _{-0.05} $ \\
-20.0 to -19.5 & -3.25 $ ^{+0.15} _{-0.22} $ & -3.16 $ ^{+0.09} _{-0.12} $ & -2.69 $ ^{+0.05} _{-0.06} $ & -2.86 $ ^{+0.07} _{-0.09} $ \\
-19.5 to -19.0 & -2.74 $ ^{+0.09} _{-0.12} $ & -2.92 $ ^{+0.08} _{-0.10} $ & -2.44 $ ^{+0.05} _{-0.06} $ & \nodata \\
-19.0 to -18.5 & -2.53 $ ^{+0.07} _{-0.08} $ & -2.62 $ ^{+0.07} _{-0.08} $ & -2.38 $ ^{+0.10} _{-0.14} $ & \nodata \\
-18.5 to -18.0 & -2.50 $ ^{+0.07} _{-0.09} $ & -2.63 $ ^{+0.08} _{-0.10} $ & \nodata & \nodata \\
-18.0 to -17.5 & -2.38 $ ^{+0.07} _{-0.09} $ & -2.70 $ ^{+0.13} _{-0.19} $ & \nodata & \nodata \\
-17.5 to -17.0 & -2.39 $ ^{+0.08} _{-0.09} $ & \nodata & \nodata & \nodata \\
-17.0 to -16.5 & -2.04 $ ^{+0.10} _{-0.13} $ & \nodata & \nodata & \nodata \\
-16.5 to -16.0 & -2.27 $ ^{+0.16} _{-0.27} $ & \nodata & \nodata & \nodata \\
\enddata

\tablecomments{\ Luminosity function data for each redshift and magnitude bin.  The numbers presented are log($\phi$).  Uncertainties do not include the effects of cosmic variance.}

\end{deluxetable}


An additional source of error is cosmic variance, in which a pencil-beam survey could be observing an over- or under-dense region of the universe.  This is accounted for using the publicly available code of \citet{trenti08}, which is based on $N$-body simulations of galaxy formation.  It takes as inputs the area of the survey, mean redshift, range of redshifts observed, the intrinsic number of detected objects, and the average incompleteness to calculate both the relative Poisson error and the relative error due to cosmic variance.  
Although cosmic variance does depend on dark matter halo mass \citep[e.g.,][]{somerville04} and thus galaxy luminosity, the cosmic variance estimates calculated using the method of \citet{trenti08} integrates over all dark matter halo masses and thus the cosmic variance estimates quoted here are average values for our sample.

These quantities were calculated for the galaxies in each
redshift bin.  The number of galaxies and completeness were chosen to be those found in the same bootstrap calculation that resulted in the chosen $\phi$.  Assuming that the Poisson error was accounted for by the bootstrap approach, the factor by which to increase the errors is given by
$\sqrt{1 + (\sigma_\text{CV} / \sigma_\text{P} )^2}$,
where $\sigma_\text{CV}$ and $\sigma_\text{P}$ are the cosmic variance and Poisson errors, respectively, found in the \citet{trenti08} code output.  This ensures that the factor is $\sim$$\sqrt{2}$ when the two error sources are of similar magnitude, and close to 1 if the cosmic variance error is negligible. 
Because cosmic variance is an uncertainly in the normalization of the luminosity function, we apply the correction to the error in $\phi^*$.
The relative importance of cosmic variance in each redshift bin is compiled in Table~\ref{tab:cv}.

\begin{deluxetable}{lc}
\tablecaption{Contribution of Cosmic Variance \label{tab:cv}}
\tablewidth{0pt}
\tabletypesize{\footnotesize}
\tablehead{
\colhead{Redshift} & Cosmic Variance
}
\startdata
$ 0.2-0.4 $ & 2.248 \\
$ 0.4-0.6 $ & 2.040 \\
$ 0.6-0.8 $ & 2.529 \\
$ 0.8-1.2 $ & 2.139 \\
\enddata

\tablecomments{Relative contribution of cosmic variance to the normalization uncertainty in each redshift bin.  The quantity displayed is $\sqrt{1 + (\sigma_\text{CV} / \sigma_\text{P} )^2}$ (see \S\ref{sec:vmax}).  The total $\phi^*$ error is calculated by increasing its uncertainty by the factor in the table.}

\end{deluxetable}

\begin{figure}[h!]
    \centering
    \includegraphics[trim = 0mm 25mm 0mm 10mm, clip=true, scale=0.6, angle=180]{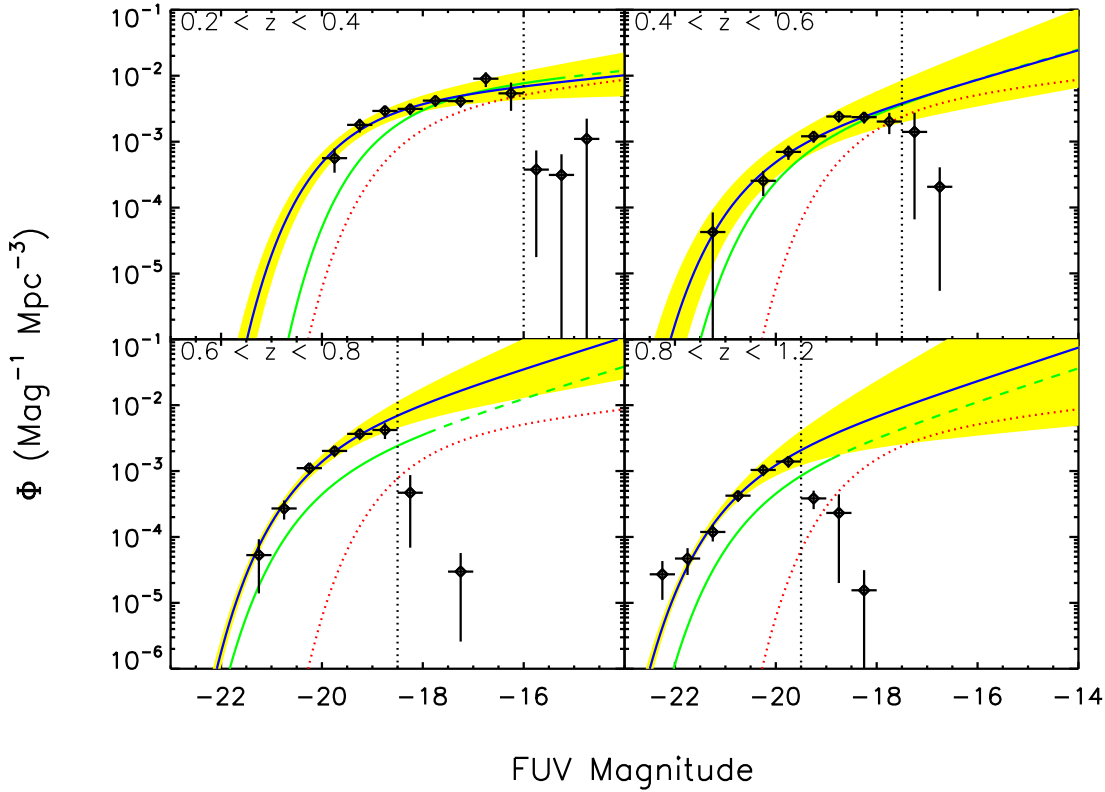}
    \caption{FUV luminosity functions for each of the four redshift bins. 
    		The $V_\text{max}$ Schechter function fit is marked with a blue line, with the $1\sigma$ error region due to $M^*$ and $\alpha$ shaded yellow.  The vertical dotted line marks the point beyond which the data are dominated by galaxies with magnitudes fainter than the completeness cutoff of 50\%.  The dotted red curve is the \citet{wyder05} LF for the local universe.  The green curve is the \citet{arnouts05} LF for each of the redshift bins shown, which becomes dashed past their respective limiting magnitudes.}
    \label{lf_boot}
\end{figure}

In each redshift bin, the data are fit with a Schechter function, given by
\begin{equation}
\phi(M) dM
=
\phi^* \left(0.4 \ln 10\right) 
10^{ 0.4 (M^*-M)(\alpha+1)}
\exp \left(
-10^{0.4(M^*-M)}
\right) dM .
\end{equation}
The free parameters are 
$\alpha$, the slope at the faint end of the LF; 
$M^*$, the magnitude at which the LF turns over;
and $\phi^*$, the density normalization.
The fit is made using MPFIT, an IDL Levenberg-Marquardt least-squares code \citep{markwardt09}.  
The data and fits are in Figure~\ref{lf_boot} and tabulated in Table~\ref{boot_params}, with errors in the Schechter parameters calculated by MPFIT.

Data points to the right of the dotted lines in Figure~\ref{lf_boot} are not included in the fit, since those magnitude bins are primarily populated by galaxies with apparent magnitudes below our 50\% completeness cutoff.  Because of this limitation, we do not put strong constraints on $\alpha$; therefore, we adopt the values and uncertainties for $\alpha$ found in \citet{arnouts05}. 
When calculating the best values for $\phi^*$ and $M^*$, $\alpha$ is fixed; its uncertainties from \citet{arnouts05} are propagated when calculating the SFRD (Section~\ref{sec:sfr}).

\begin{deluxetable}{lllrll}
\tablecaption{$V_\text{max}$ Schechter Function Parameters \label{boot_params}}
\tablewidth{0pt}
\tabletypesize{\footnotesize}
\tablehead{
\colhead{Redshift} & \colhead{$\phi^*/10^{-3}$\tablenotemark{\dagger}} & \colhead{$M^*$} & \colhead{$\alpha$\tablenotemark{\ddagger}} & \colhead{$\rho/10^{26}$} & \colhead{SFR Density$/10^{-2}$} \\
\colhead{} & \colhead{(Mpc$^{-3}$)} & \colhead{(AB mag)} & \colhead{} & \colhead{(erg/s/Hz/Mpc$^3$)} & \colhead{(M$_\odot$/yr/Mpc$^3$)}
}
\startdata
$ 0.2-0.4 $ &  $ 4.45 \pm 1.62 $  &  $ -19.24 \pm 0.23 $ &  $ -1.19 \pm 0.15 $ &  $ 1.092 \pm 0.419 $ &  $ 0.963 \pm 0.369 $ \\ 
$ 0.4-0.6 $ &  $ 1.18 \pm 0.88 $  &  $ -20.14 \pm 0.39 $ &  $ -1.55 \pm 0.21 $ &  $ 1.124 \pm 0.863 $ &  $ 0.992 \pm 0.761 $ \\ 
$ 0.6-0.8 $ &  $ 4.38 \pm 2.41 $  &  $ -19.95 \pm 0.15 $ &  $ -1.60 \pm 0.26 $ &  $ 4.177 \pm 2.359 $ &  $ 3.685 \pm 2.081 $ \\ 
$ 0.8-1.2 $ &  $ 1.87 \pm 1.34 $  &  $ -20.50 \pm 0.21 $ &  $ -1.63 \pm 0.45 $ &  $ 3.059 \pm 2.384 $ &  $ 2.698 \pm 2.103 $ \\ 
\enddata


\tablenotetext{\dagger}{The $\phi^*$ uncertainties include the contribution of cosmic variance (Table~\ref{tab:cv}).}
\tablenotetext{\ddagger}{Values and uncertainties for $\alpha$ are taken from \citet{arnouts05}.}

\end{deluxetable}



\subsection{MLE Method} \label{sec:mle}

The second method for determining the best Schechter function parameters has the advantage of not needing to bin the data.  For clarity, the equations presented in this section are in terms of luminosity rather than magnitude.
We follow the MLE procedure derived in \citet{ciardullo13}, in which the relative probability $P$ of a given function fitting the data is
\begin{equation}
\ln P  
=
- \int_{z_1}^{z_2} \int_{L_\text{min}(z)}^\infty
\phi'(L) \ dL \ dV 
+
\sum_i^N \ln \phi'(L_i), 
\end{equation}
where $z_1$ to $z_2$ defines the redshift bin,
$L_\text{min}$ is the faintest luminosity that can be observed at the given redshift, 
$\phi'(L)$ is the luminosity function modified by any selection effects (including incompleteness), 
and $L_i$ is the luminosity of a given galaxy.  The specific value of $\ln P$ in unimportant; it is only used for comparing across different model parameters.

The integrals are by necessity integrated numerically; the Schechter function alone can be integrated analytically, but for this likelihood formulation, a completeness term must be included.  We evaluate $\ln P$ for a range of $M^*$ and $\phi^*$ values.  
As found in Section~\ref{sec:vmax}, our data do not go deep enough to constrain $\alpha$, so we set $\alpha$ to those found by \citet{arnouts05}, and use the $\alpha$ uncertainties when calculating the SFRD.)
We also exclude galaxies with magnitudes fainter than the fitting cutoff used in Section~\ref{sec:vmax}.
To implement this, we use a proxy for $\phi^*$, since the value for $\phi^*$ is strongly dependent upon the values of $M^*$ and $\alpha$.  This proxy, referred to as $\phi_\text{tot}$, is defined as
\begin{equation}
\phi_\text{tot} = \int_{L_\text{min}}^\infty
\phi(L) \ dL , 
\end{equation}
where $L_\text{min}$ is the detection limit of the given redshift bin and $\phi(L)$ is the Schechter function.
It represents the approximate volume density of galaxies above $L_\text{min}$.  Unlike $\phi^*$, $\phi_\text{tot}$ doesn't change significantly with $M^*$ or $\alpha$.  Therefore, when searching through a grid of Schechter parameters, we make an evenly-spaced grid of $\phi_\text{tot}$ values, which we translate into a $\phi^*$ before calculating each likelihood.

\begin{figure}[h!]
    \centering
    \includegraphics[trim = 0mm 25mm 0mm 10mm, clip=true, scale=0.6, angle=180]{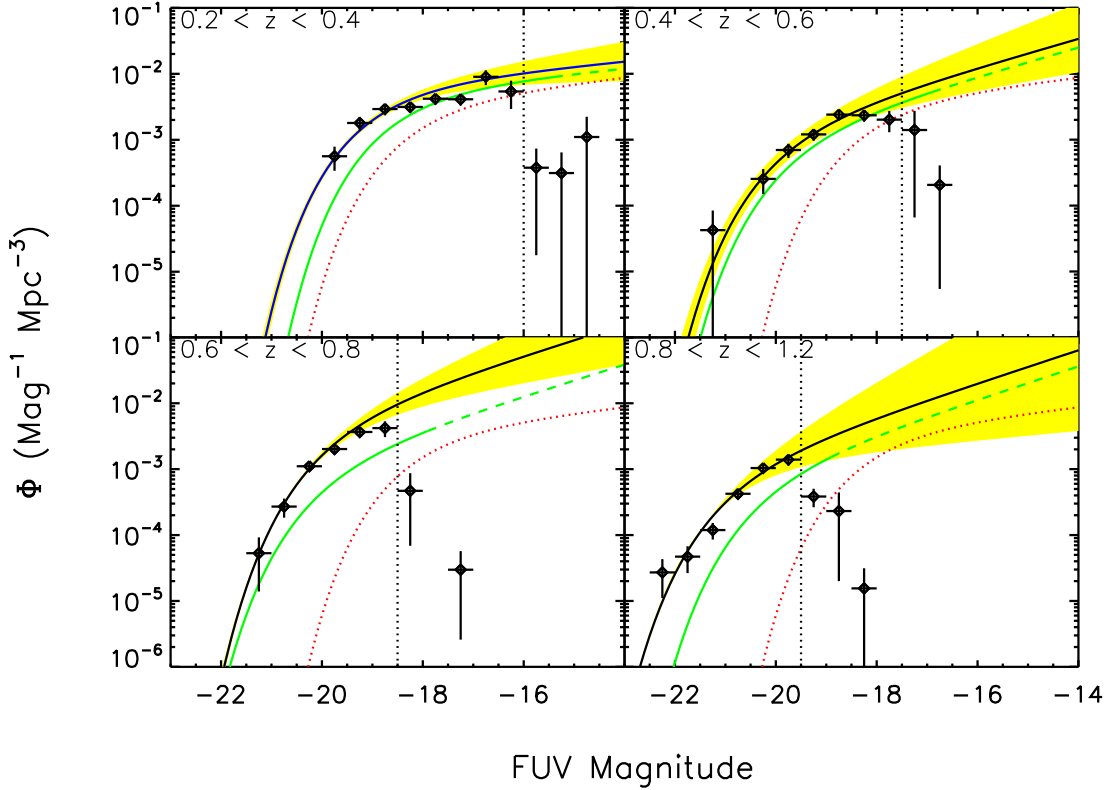}
    \caption{Same as Figure~\ref{lf_boot}, but using the MLE Schechter function fitting method.  The binned data points are included for reference, but were not used in the fitting process.}
    \label{lf_mle}
\end{figure}

\begin{figure}[h!]
    \centering
    \includegraphics[trim = 20mm 20mm 35mm 25mm, clip=true, scale=0.23, angle=180, page=2]{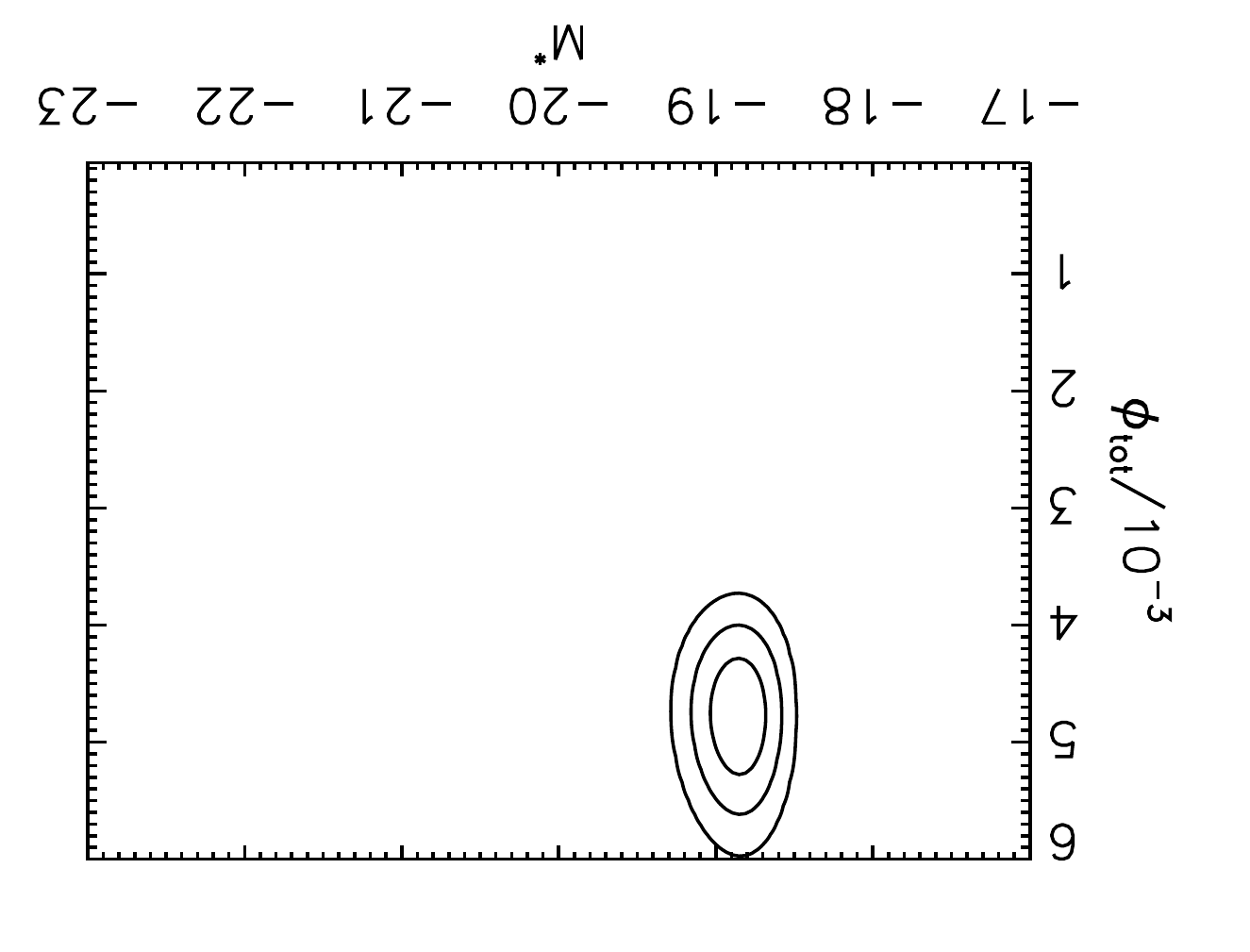}
    \includegraphics[trim = 20mm 20mm 20mm 25mm, clip=true, scale=0.23, angle=180, page=3]{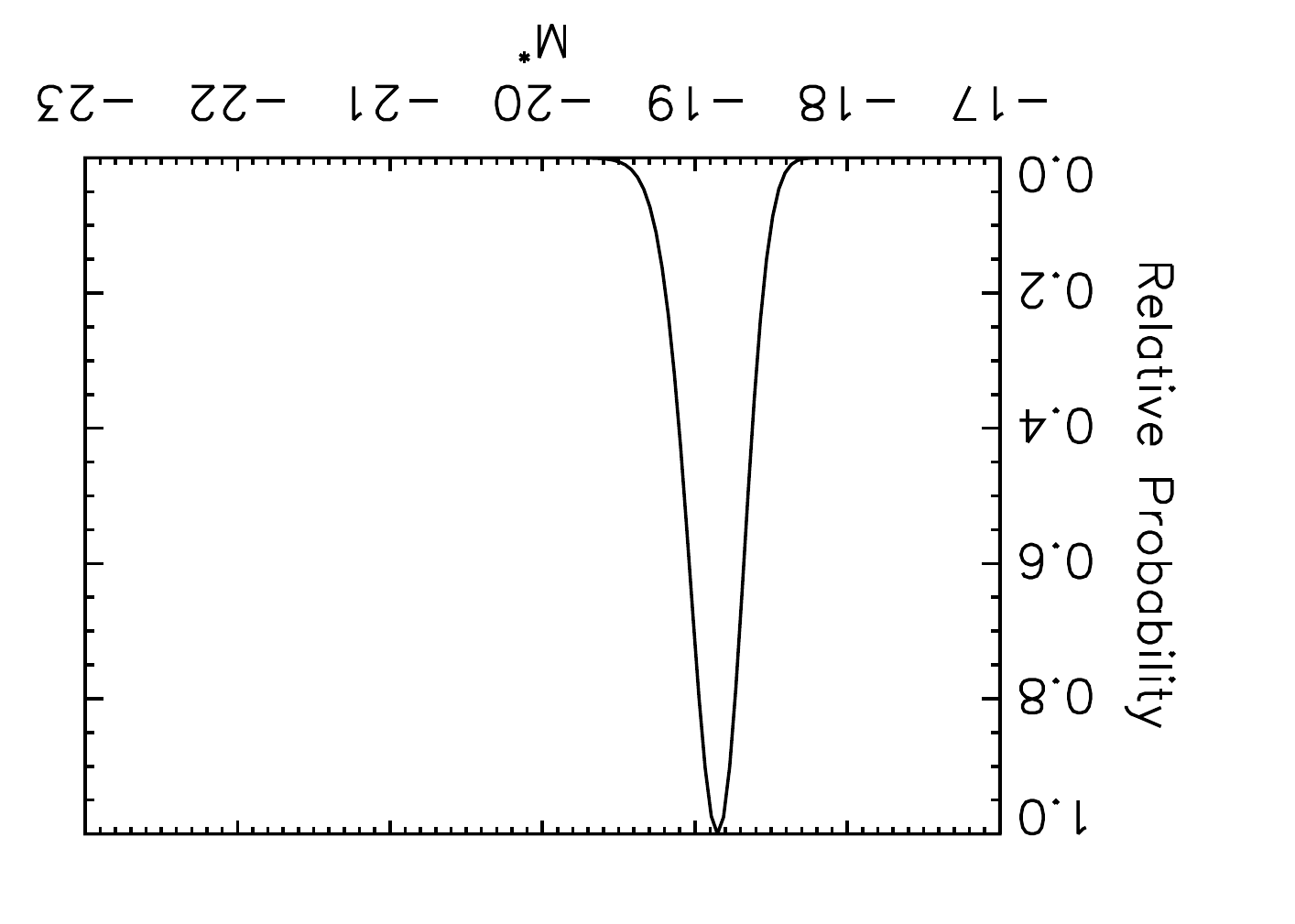}
    \includegraphics[trim = 20mm 20mm 20mm 25mm, clip=true, scale=0.23, angle=180, page=4]{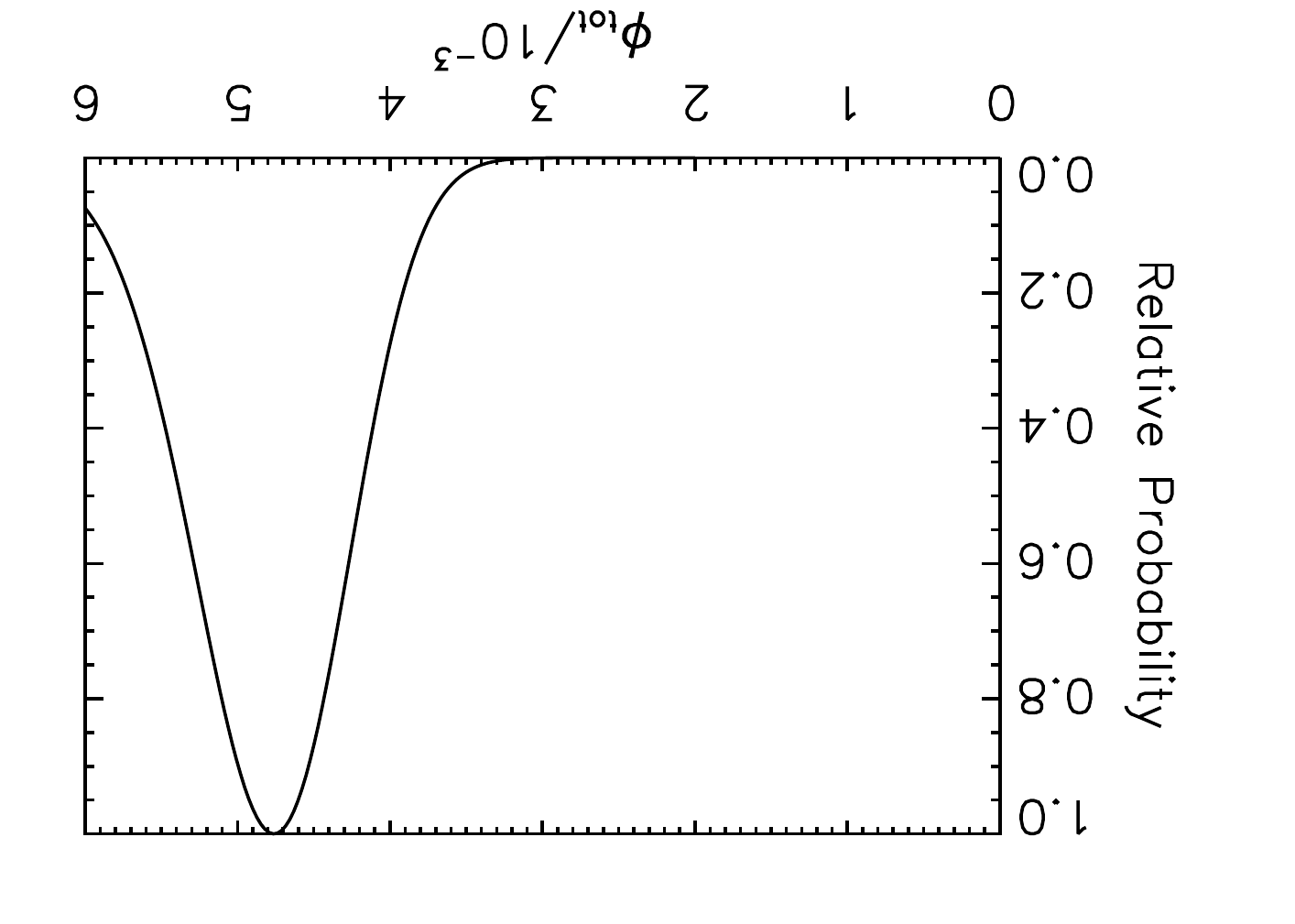}
    \includegraphics[trim = 20mm 20mm 35mm 25mm, clip=true, scale=0.23, angle=180, page=6]{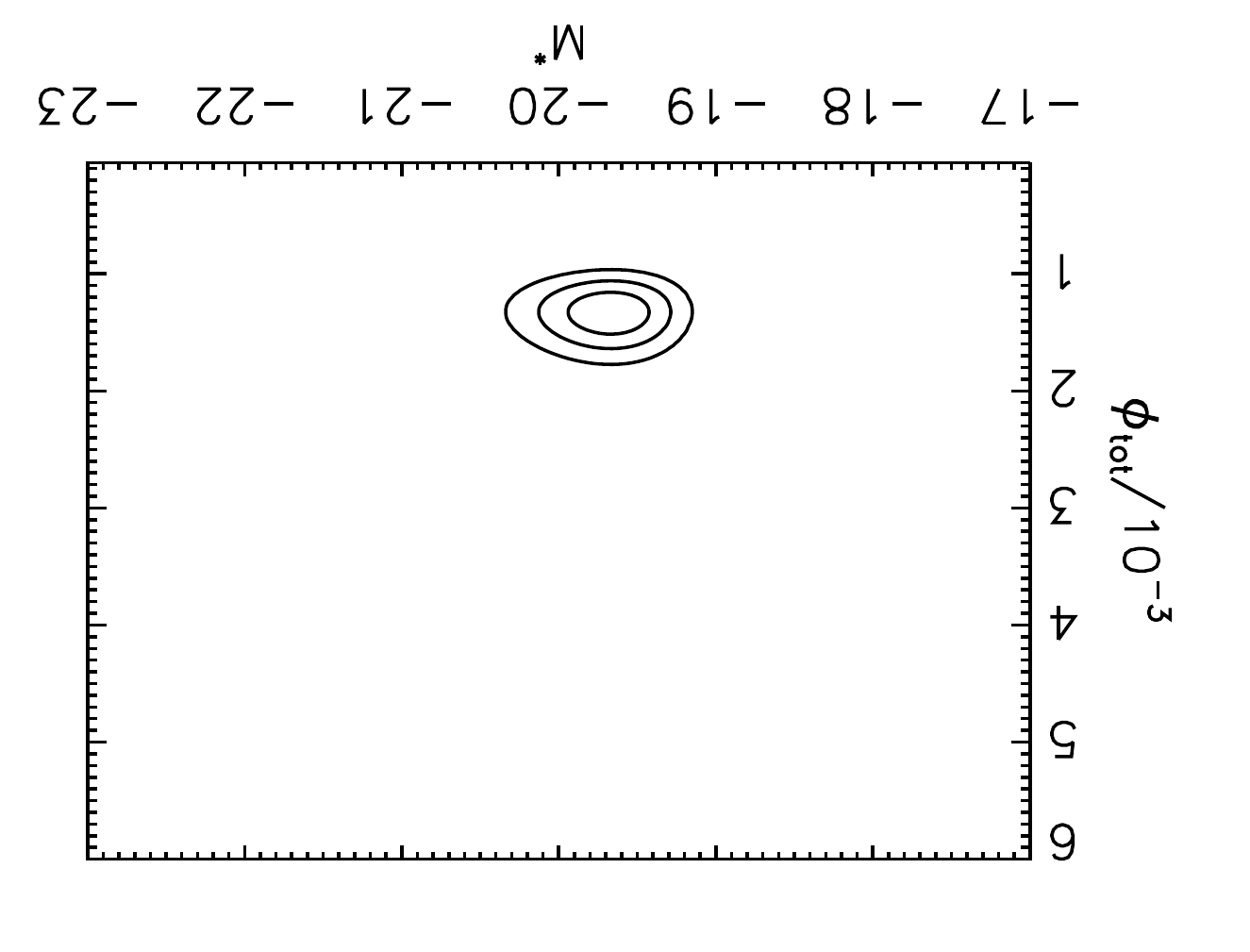}
    \includegraphics[trim = 20mm 20mm 20mm 25mm, clip=true, scale=0.23, angle=180, page=7]{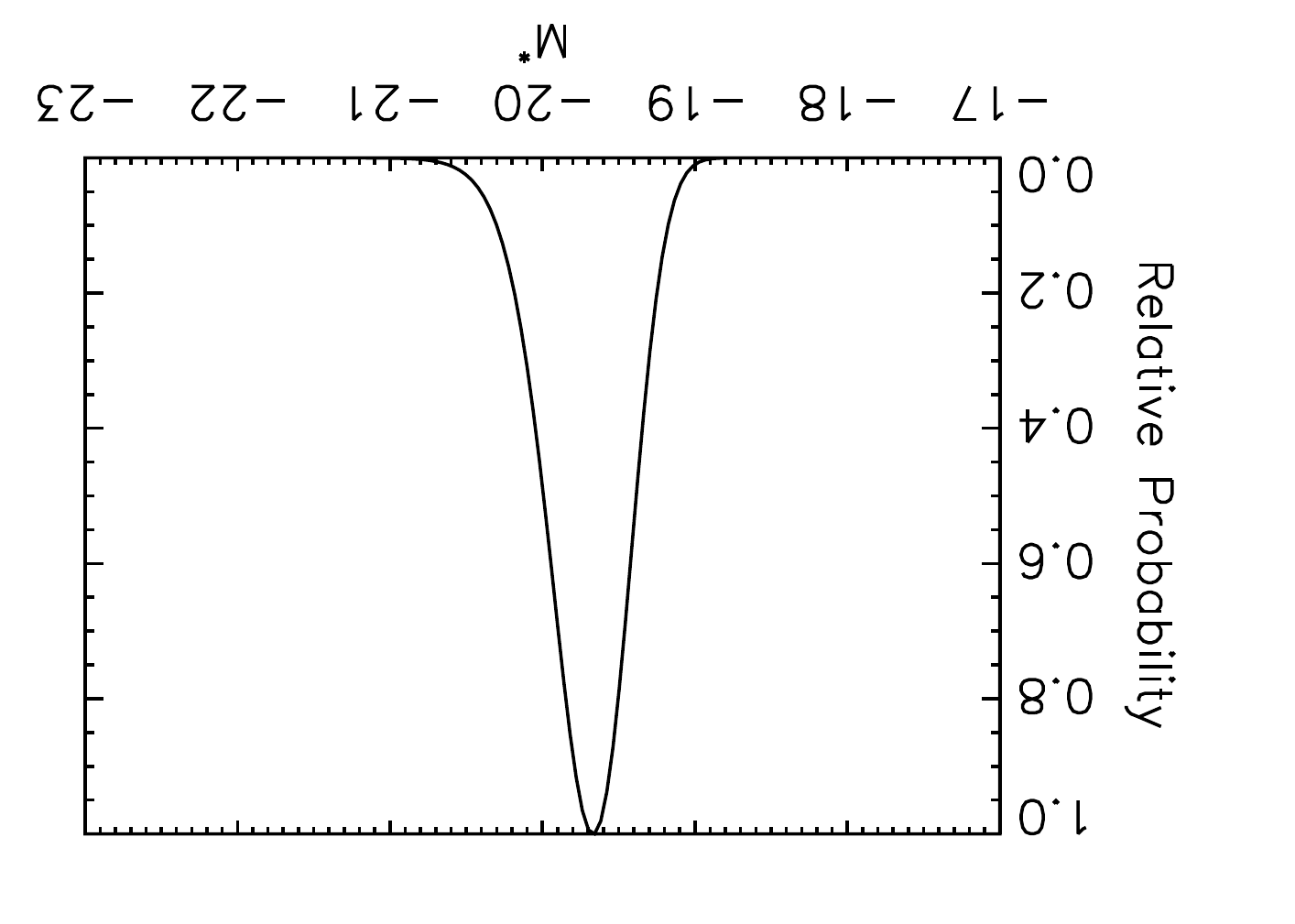}
    \includegraphics[trim = 20mm 20mm 20mm 25mm, clip=true, scale=0.23, angle=180, page=8]{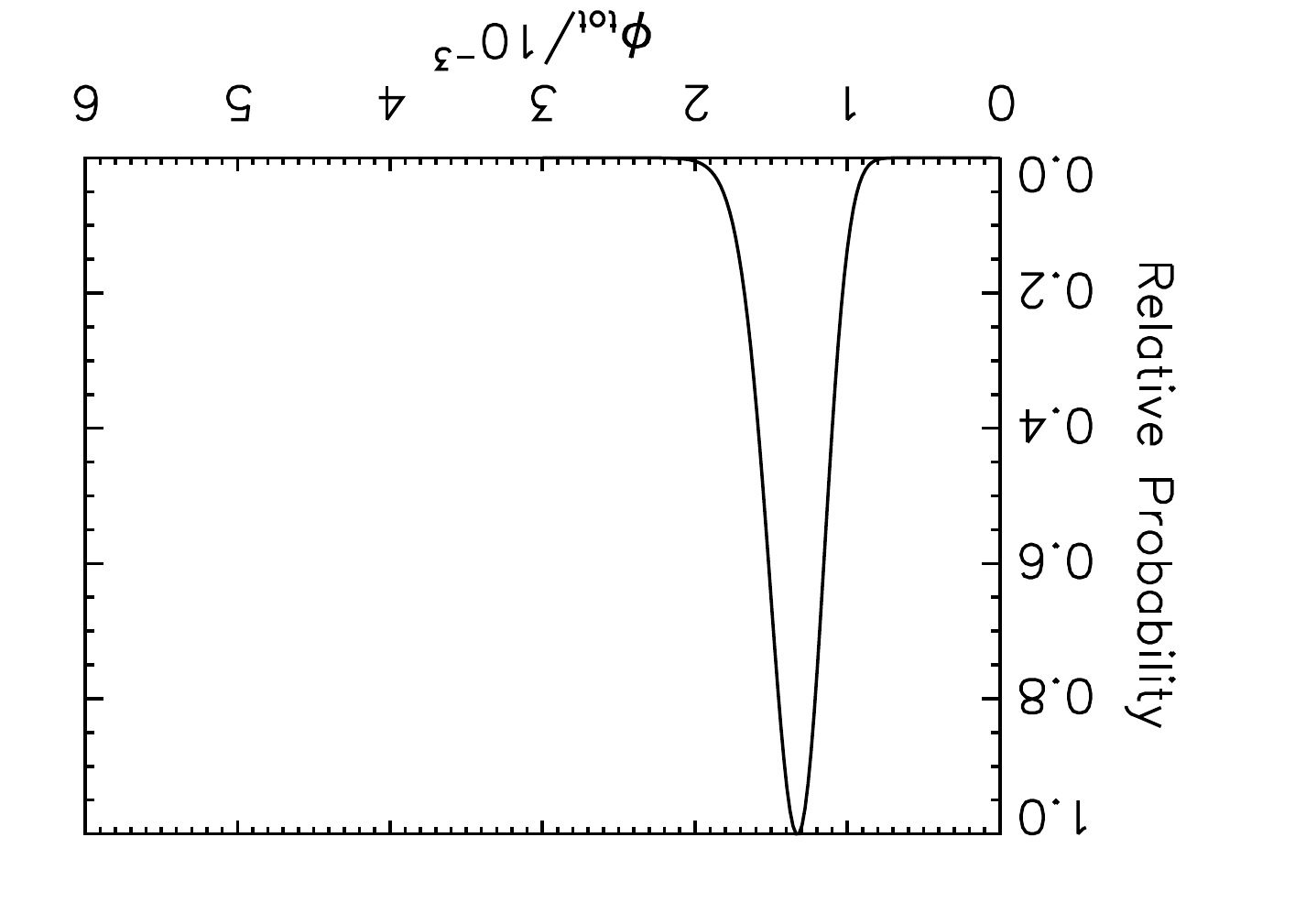}
    \includegraphics[trim = 20mm 20mm 35mm 25mm, clip=true, scale=0.23, angle=180, page=10]{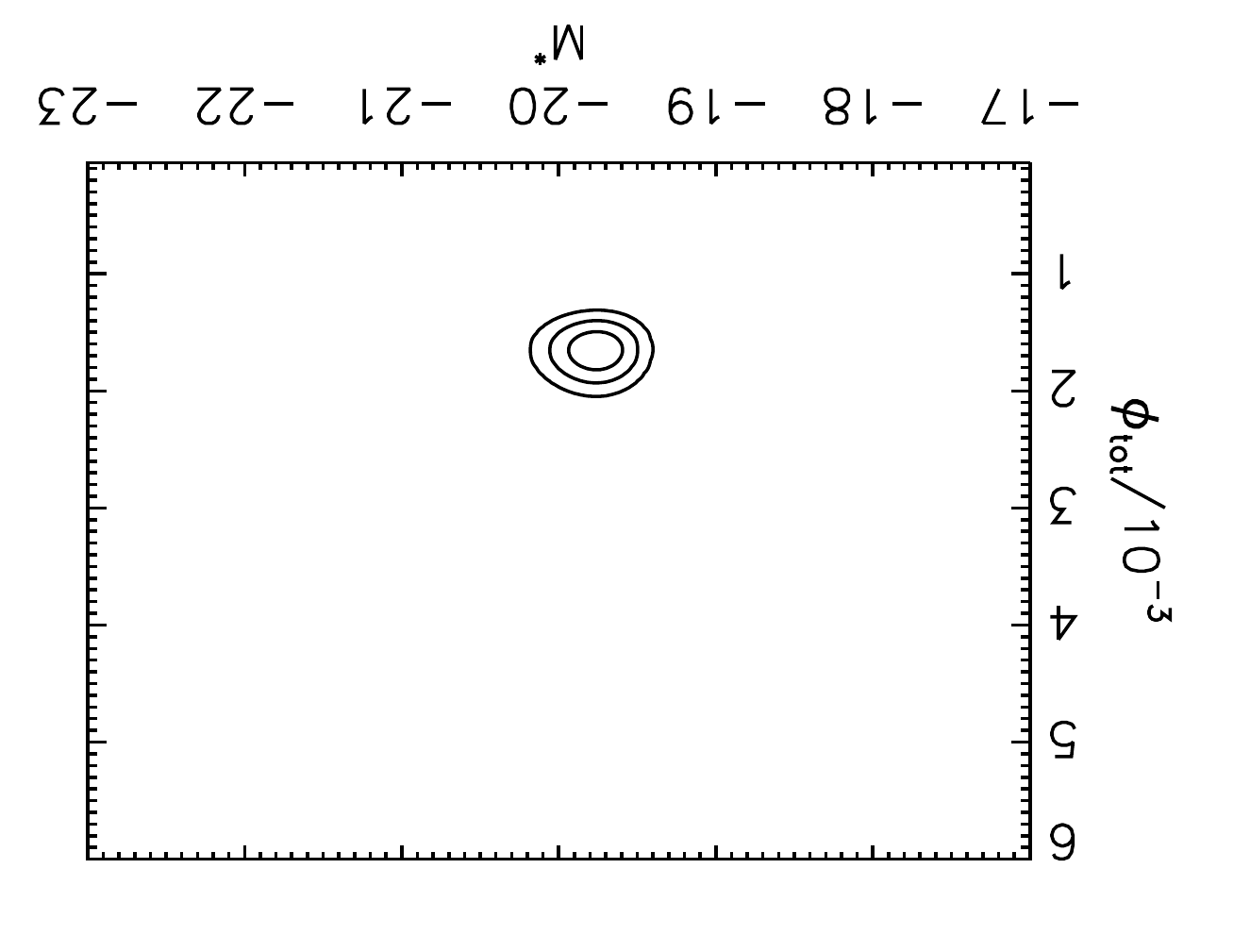}
    \includegraphics[trim = 20mm 20mm 20mm 25mm, clip=true, scale=0.23, angle=180, page=11]{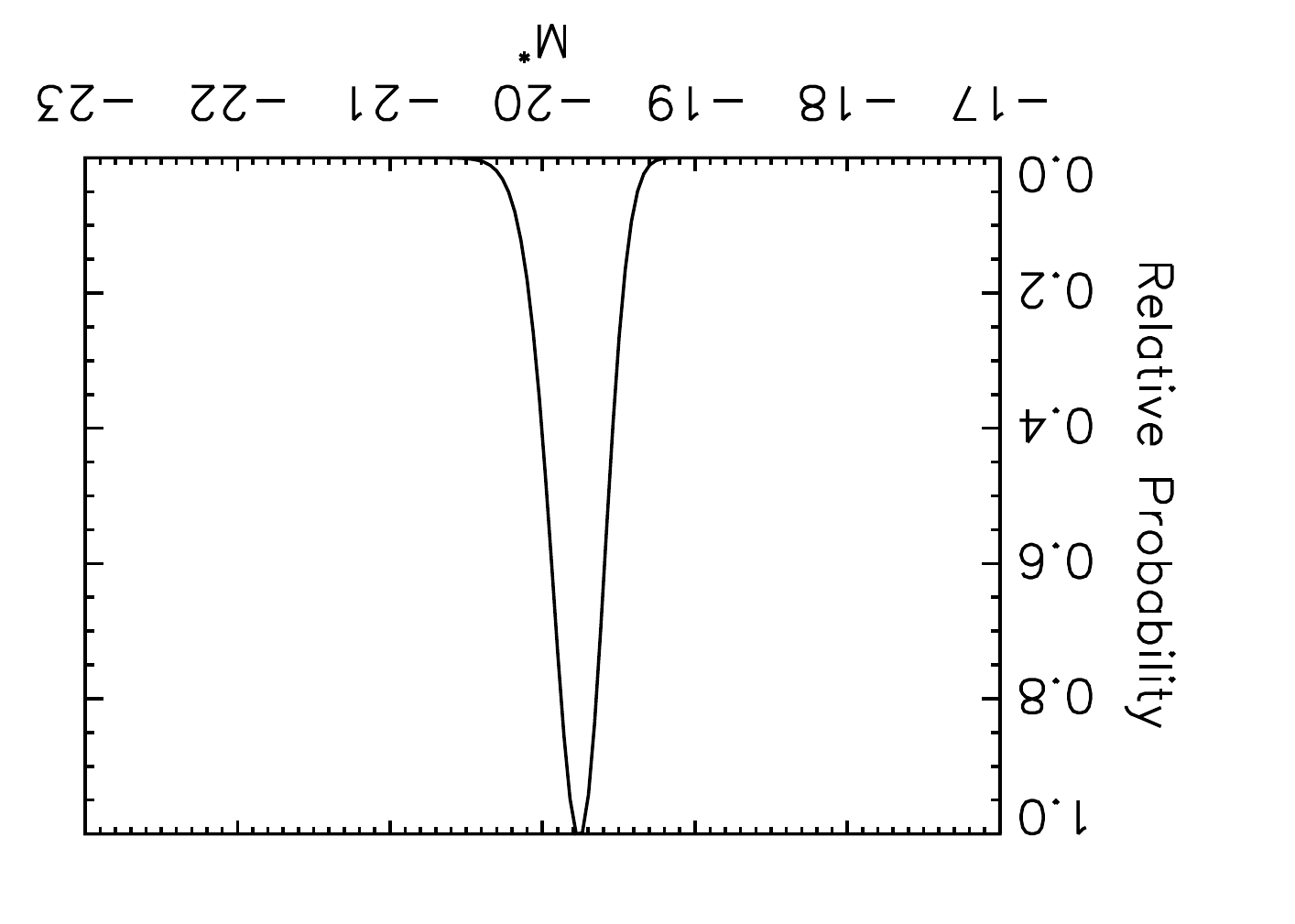}
    \includegraphics[trim = 20mm 20mm 20mm 25mm, clip=true, scale=0.23, angle=180, page=12]{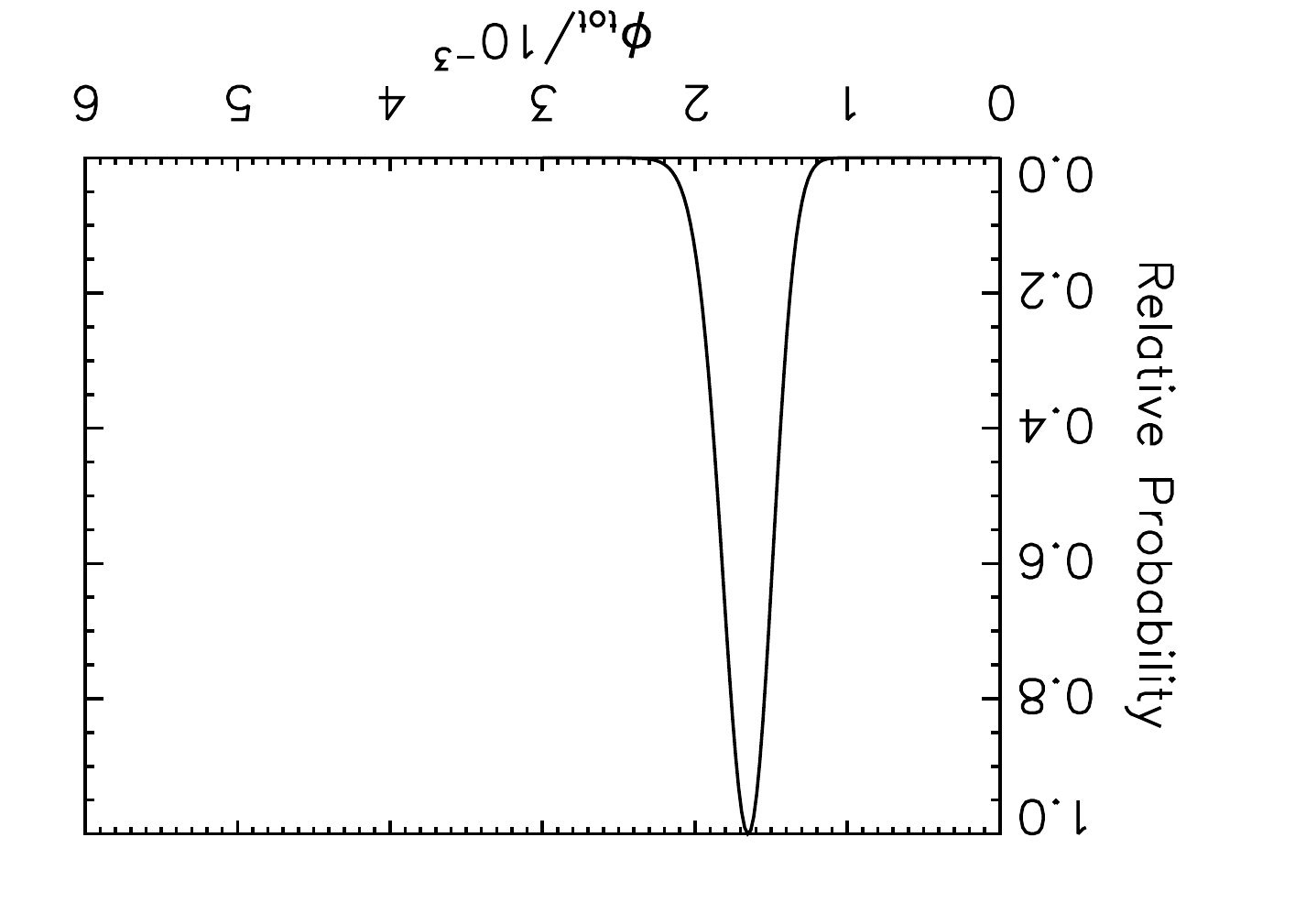}
    \includegraphics[trim = 20mm 20mm 35mm 25mm, clip=true, scale=0.23, angle=180, page=14]{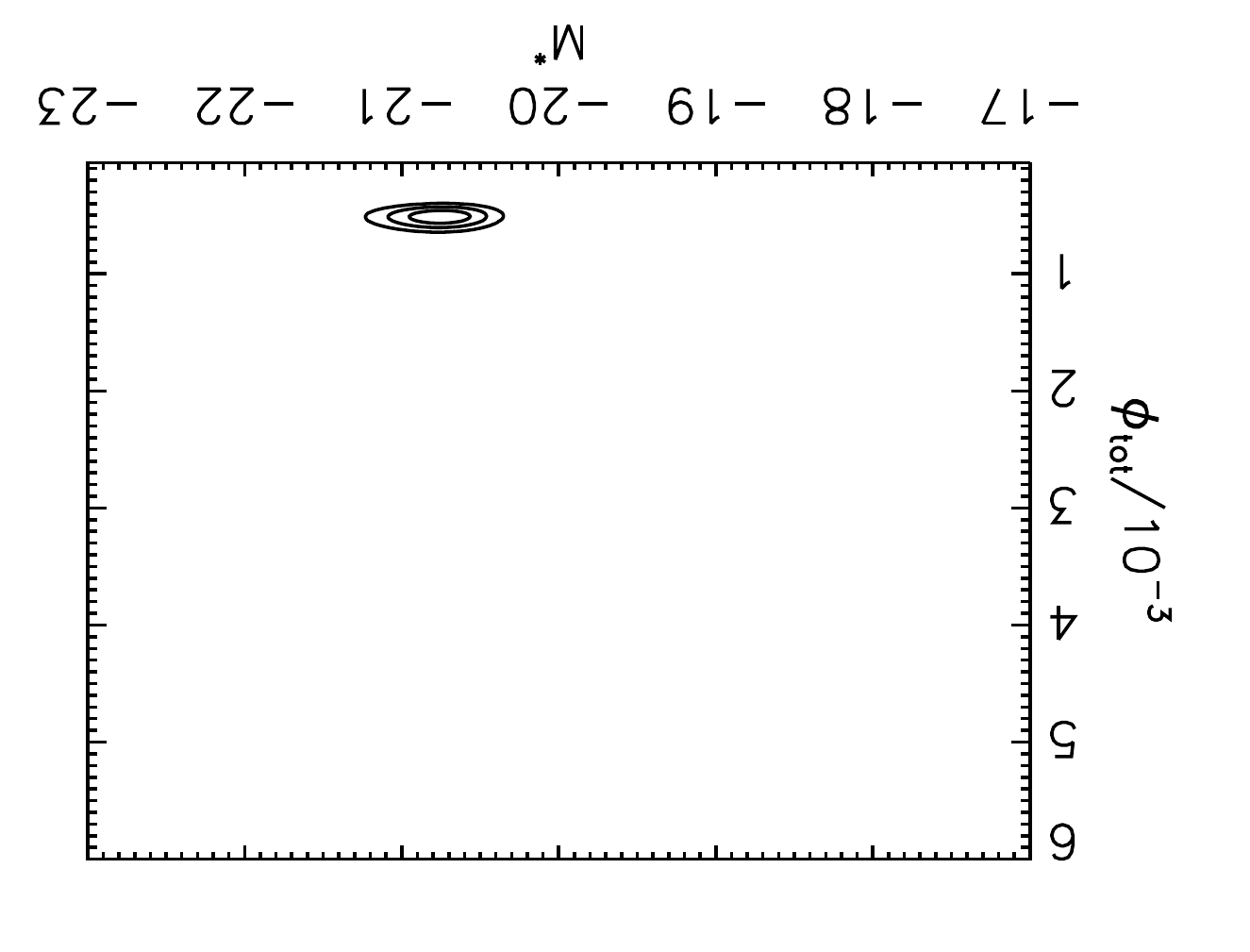}
    \includegraphics[trim = 20mm 20mm 20mm 25mm, clip=true, scale=0.23, angle=180, page=15]{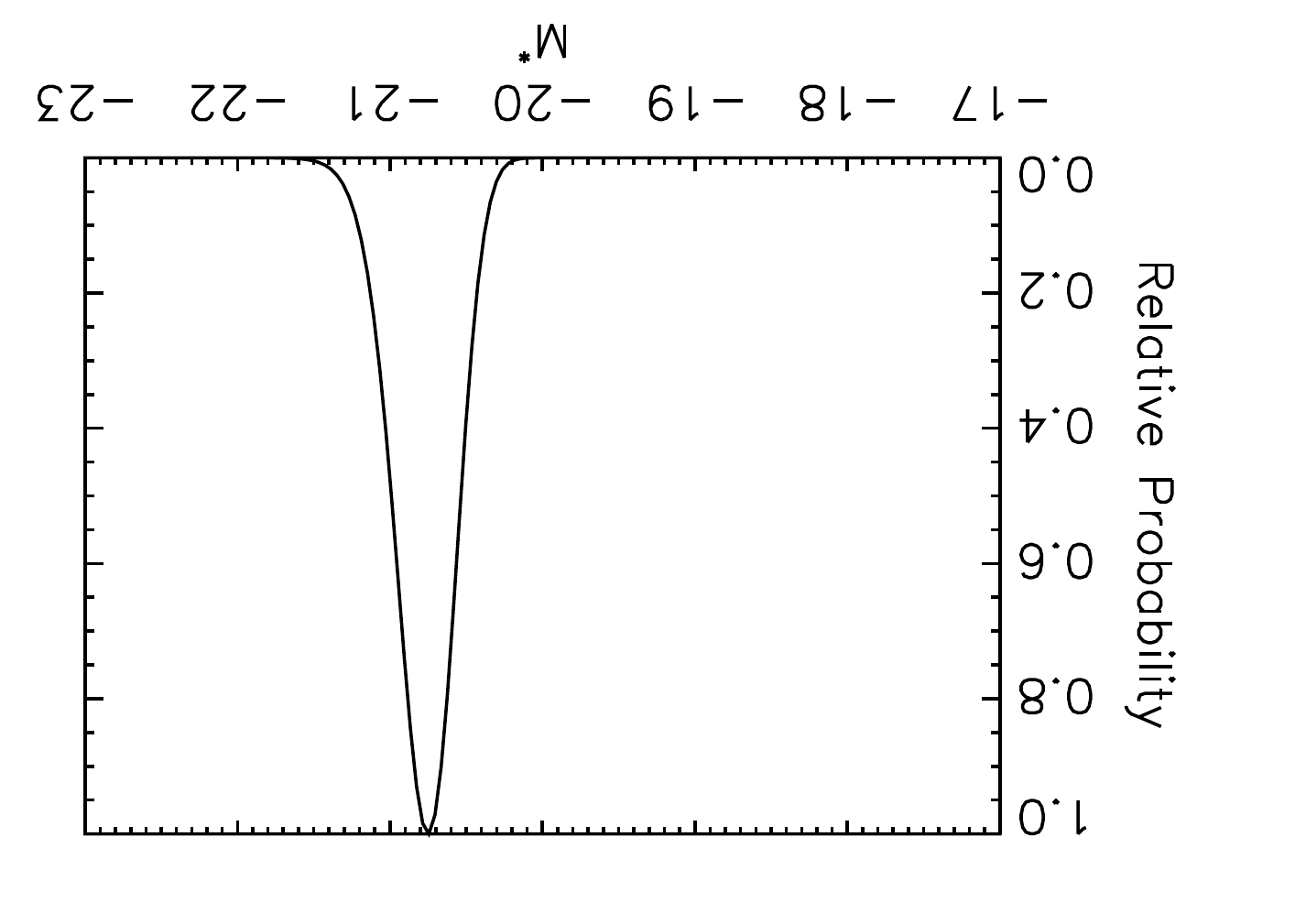}
    \includegraphics[trim = 20mm 20mm 20mm 25mm, clip=true, scale=0.23, angle=180, page=16]{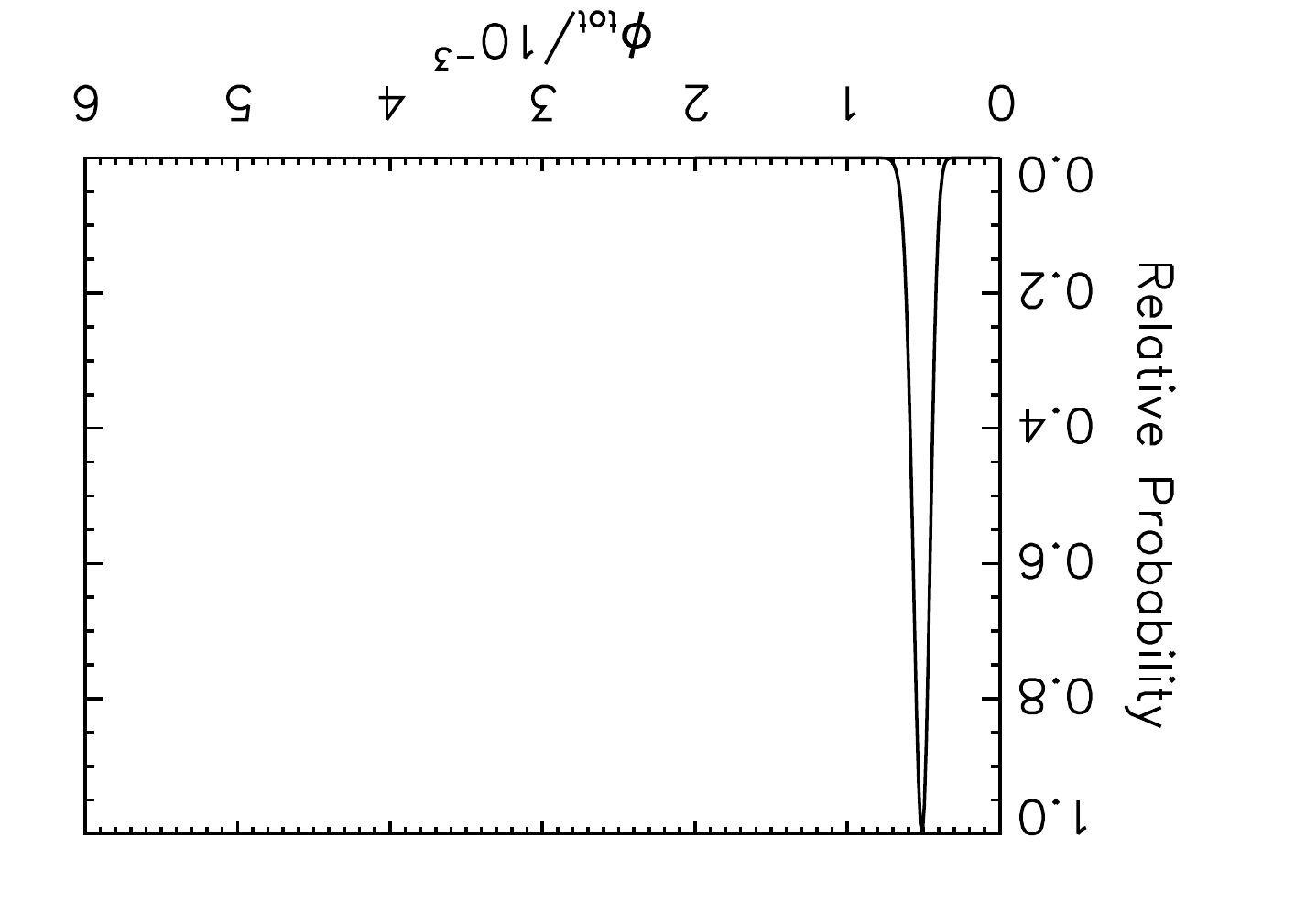}
    \caption{Schechter function parameters from the MLE fitting method, assuming a fixed $\alpha$ from \citet{arnouts05}.  Each row is a redshift bin ($z=0.2-0.4$ at top, $z=0.8-1.2$ at bottom).  The first column shows the $1\sigma$, $2\sigma$, and $3\sigma$ best-fit contours.  The second and third columns are the relative probability distributions of $M^*$ and $\phi_\text{tot}$, respectively.}
    \label{fig:mle}
\end{figure}

The results of our fitting are shown in Figure~\ref{lf_mle}.
Details about the best-fit parameters are in Figure~\ref{fig:mle}, which is divided into three parts.  The first column shows the two-dimensional distribution of log likelihoods for each redshift bin.  The second and third columns are the resulting probability distributions of $M^*$ and $\phi_\text{tot}$, respectively.  
The highest likelihood parameter values from these distributions are listed in Table~\ref{mle_params}, in which $\phi^*$ has been derived from $\phi_\text{tot}$ using the best-fit $M^*$.

The two-dimensional likelihoods confirm that there are no fitting degeneracies, which is information that can only be found with a technique that calculates likelihoods for a whole grid of variables.  Had we been fitting for all three Schechter function parameters, however, it is likely that there would be strong degeneracies.  In addition, this method shows that the $M^*$ and $\phi_\text{tot}$ probability distributions can be treated as Gaussian.

\begin{deluxetable}{lllrll}
\tablecaption{MLE Schechter Function Parameters \label{mle_params}}
\tablewidth{0pt}
\tabletypesize{\footnotesize}
\tablehead{
\colhead{Redshift} & \colhead{$\phi^*/10^{-3}$\tablenotemark{\dagger}} & \colhead{$M^*$} & \colhead{$\alpha$\tablenotemark{\ddagger}} & \colhead{$\rho/10^{26}$} & \colhead{SFR Density$/10^{-2}$} \\
\colhead{} & \colhead{(Mpc$^{-3}$)} & \colhead{(AB mag)} & \colhead{} & \colhead{(erg/s/Hz/Mpc$^3$)} & \colhead{(M$_\odot$/yr/Mpc$^3$)}
}
\startdata
$ 0.2-0.4 $ &  $ 6.81 \pm 1.42 $  &  $ -18.85 \pm 0.12 $ &  $ -1.19 \pm 0.15 $ &  $ 1.203 \pm 0.293 $ &  $ 1.061 \pm 0.258 $ \\ 
$ 0.4-0.6 $ &  $ 2.23 \pm 0.47 $  &  $ -19.66 \pm 0.20 $ &  $ -1.55 \pm 0.21 $ &  $ 1.399 \pm 0.534 $ &  $ 1.234 \pm 0.471 $ \\ 
$ 0.6-0.8 $ &  $ 6.65 \pm 1.21 $  &  $ -19.78 \pm 0.10 $ &  $ -1.60 \pm 0.26 $ &  $ 5.463 \pm 2.300 $ &  $ 4.818 \pm 2.028 $ \\ 
$ 0.8-1.2 $ &  $ 1.36 \pm 0.19 $  &  $ -20.74 \pm 0.12 $ &  $ -1.63 \pm 0.45 $ &  $ 2.980 \pm 1.646 $ &  $ 2.629 \pm 1.452 $ \\ 
\enddata


\tablenotetext{\dagger}{The $\phi^*$ uncertainties include the contribution of cosmic variance (Table~\ref{tab:cv}).}
\tablenotetext{\ddagger}{Values and uncertainties for $\alpha$ are taken from \citet{arnouts05}.}

\end{deluxetable}



\section{Star Formation Rate Density} \label{sec:sfr}

\begin{figure}[h!]
    \centering
    \includegraphics[trim = 0mm 80mm 0mm 30mm, clip=true, scale=0.6]{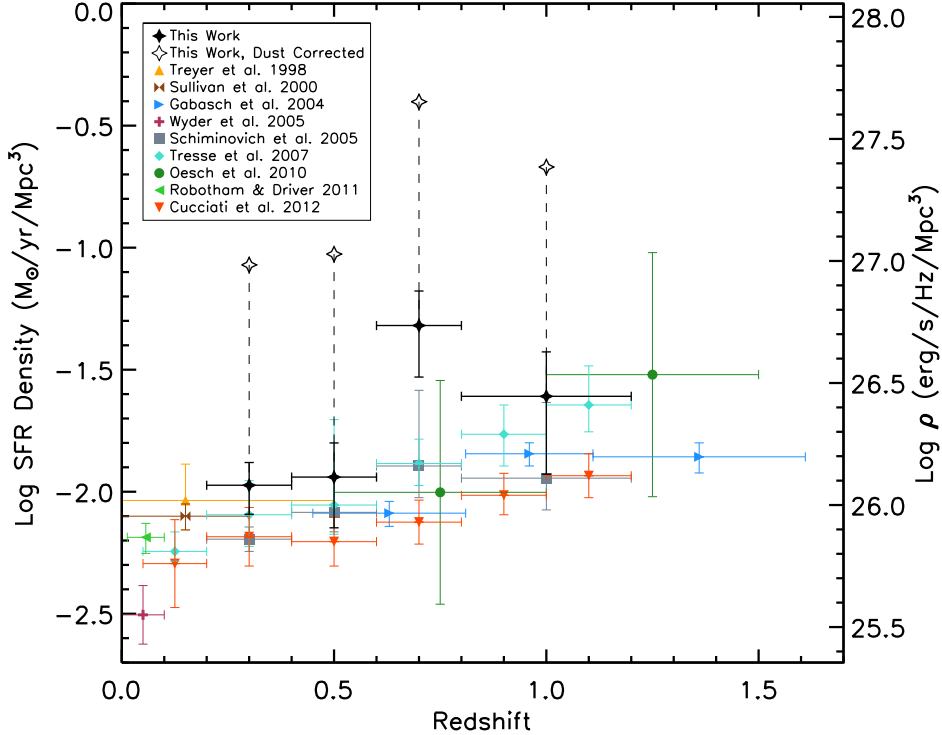}
    \caption{SFR density and luminosity density for each redshift bin, compared to literature values.  Data comes from the MLE fits with fixed $\alpha$, and is shown both with and without a dust correction.  Uncertainties include the contribution from cosmic variance.  The large SFR density at $z=0.7$ is due to the galaxy overdensity at that redshift (see Figure~\ref{z_dist}).
    \textit{List of references:} 
    \citet{treyer98} (rest-frame 2000~\AA, data from the FOCA balloon-borne UV camera, WIYN, and William Herschel Telescope); 
    \citet{sullivan00} (same as \citet{treyer98}, but with larger field of view); 
    \citet{gabasch04} (rest-frame 1500~\AA, data from FORS Deep Field on VLT and NTT); 
    \citet{wyder05} (rest-frame 1500~\AA, data from \textit{GALEX});
    \citet{schiminovich05} (rest-frame 1500~\AA, data from \textit{GALEX});
    \citet{tresse07} (rest-frame 1500~\AA, data from VLT);
    \citet{oesch10} (rest-frame 1500~\AA, data from HST);
    \citet{robotham11} (rest-frame 1500~\AA, data from \textit{GALEX}); 
    \citet{cucciati12} (rest-frame 1500~\AA, data from VLT).  }
    \label{sfr_ltot_u}
\end{figure}

Integrating the luminosity function gives the luminosity density (the FUV luminosity per unit comoving volume), which can then be converted into a SFR density.
To calculate the luminosity density, we use the Schechter function fit parameters in an analytical formula from \citet{gallego95},
\begin{equation}
\rho
=
\int_0^\infty L \ \phi(L) \ dL
=
\phi^* L^* \Gamma(2+\alpha) .
\end{equation}
The resulting luminosities per comoving volume are tabulated in Tables \ref{boot_params} and \ref{mle_params}.  The MLE-derived luminosity densities are plotted in Figure \ref{sfr_ltot_u} along with several literature values across a similar redshift range.  For uniformity, these literature values were derived from rest-frame FUV data, and they were corrected to our assumed flat $\Lambda$CDM cosmology as needed.

We then calculate the SFR density as a function of redshift.  We chose the UV SFR conversion from \citet{hao11}, which is valid for normal star-forming galaxies.  
It assumes a constant SF history and uses a Kroupa initial mass function \citep{kroupa03} with masses from 0.1~$\msun$ to 100~$\msun$.
It is expressed as
\begin{equation}
\text{SFR} = 8.82 \times 10^{-29} \ L_\text{FUV} \ ,
\end{equation}
where the SFR is measured in $M_\odot$/yr and $L_\text{FUV}$ is the rest-frame FUV luminosity, measured in erg/s/Hz.  Using our luminosity density, we calculate the SFR density for each redshift bin, also listed in Tables \ref{boot_params} and \ref{mle_params}.  
Cosmic variance, as listed in Table~\ref{tab:cv}, is included in the $\rho$ and SFRD uncertainties.

The MLE-derived SFR densities are plotted with literature values in Figure \ref{sfr_ltot_u}.  
The $V_\text{max}$-derived SFR densities and uncertainties are very similar to those found using the MLE method.
For comparison, we individually calculated the SFRs for the literature data shown, either by converting their published luminosity densities to a SFRD or from modifying their stated SFR law.
Our results are in good agreement with these literature values, except at $z=0.7$, which is the known galaxy overdensity.

\section{Conclusion} \label{sec:con}

We have used \swift\ UVOT data of the CDF-S to calculate FUV luminosity functions and star formation rate densities for $z=0.2-0.4$, $0.4-0.6$, $0.6-0.8$, and $0.8-1.2$.  We used two updated techniques to measure the LFs.  The first of these was the traditional $V_\text{max}$ method combined with a bootstrap for reliable uncertainties, which is an improvement upon the standard $V_\text{max}$ procedure.  The second of these used an MLE method to calculate the probability distribution for each of the LF Schechter fitting parameters.  We find that using either technique, our data do not strongly constrain the faint-end slope of the LF, $\alpha$.  They do, however, yield values for the luminosity and SFR densities that are consistent with the literature.

It is worthwhile to compare our method and results to those of \citet{arnouts05} and \citet{schiminovich05}, which use \textit{GALEX} observations in a similar manner to measure the luminosity functions and SFR densities.  Although the \textit{GALEX} observations cover an area $\sim$10 times larger and go $\sim$1~mag deeper than our UVOT survey,  the number of identified galaxies is remarkably similar: the \textit{GALEX} work utilizes 1039, and here we use 730.  This demonstrates the utility of UVOT's higher resolution for this type of study.   The resulting measurements of the SFRD have uncertainties that are about five times larger, though half of that difference can be attributed to our addition of cosmic variance as a source of error.

Comparing our FUV-derived SFR densities to literature values over $0 < z \lesssim 1.5$ (Figure~\ref{sfr_ltot_u}), we find that our results are broadly similar.  The only substantial difference is at $z=0.7$, which is due to a known CDF-S galaxy overdensity.
Without including this extreme data point, we find that the SFRD evolves as $(1+z)^n$ with $n=1.88 \pm 1.32$, which is consistent with $n=2.5 \pm 0.7$ found by \citet{schiminovich05} over the same redshift range.
This range of SFR densities at each redshift may be pointing to the as yet unknown spread due to cosmic variance \citep{madau14}.  The addition of UVOT data from the CDF-S is a critical piece to understanding this component of the universe's star formation history.

An additional difficulty when using rest-frame UV data is determining how to properly account for dust extinction.  There are many possible dust extinction curves to use \citep[i.e.,][]{cardelli89, misselt99, charlot00, calzetti00, gordon03}, which each have different slopes ($R_V$) and different strengths of the 2175~\AA\ dust bump.  Recent work suggests that the extinction curve changes from galaxy to galaxy and even changes within a given galaxy, so that broadly applying a single well-determined curve is still problematic \citep[e.g.,][]{hoversten11, kriek13}. 
As seen in Table~\ref{dust_corr_table} and Figure~\ref{sfr_ltot_u}, the FUV attenuation correction is quite substantial: the correction to the SFR density is $\sim$1~dex.  Even a small uncertainty in the extinction law can make a large difference in the estimated attenuation.  For this reason, we have chosen to compare our results to the observed (rather than dust corrected) SFRDs from the literature in Figure~\ref{sfr_ltot_u}.  

Our results for the evolution in the rest-frame FUV LF and SFRD over the redshift range $0.2 < z < 1.2$, while consistent with other FUV estimates in the literature, highlight the effects of cosmic variance in our estimates of the evolution of the SFRD with cosmic time.  Observations of multiple fields are required to provide a robust estimate of the evolution of the SFRD with redshift.  Our observations with the four NUV filters on \swift\ UVOT provide well-constrained rest-frame ultraviolet spectral energy distributions in the ultraviolet from which to extract FUV magnitudes used to determine both the SFR and extinction.  We plan to obtain similarly deep UVOT observations in several other deep multi-wavelength fields in the near future with which will help provide stronger constraints on
the estimates of SFRD out to $z\sim1$.


\acknowledgements
We are grateful to the anonymous referee for numerous helpful comments and suggestions.
We acknowledge support from NASA Astrophysics Data Analysis grant NNX09AC87G. 
The Institute for Gravitation and the Cosmos is supported by the Eberly College of Science and the Office of the Senior Vice President for Research at the Pennsylvania State University. 
This work is sponsored at PSU by NASA contract NAS5-00136 and at MSSL by funding from the Science and Technology Facilities Council (STFC). 
This research has made use of NASA's Astrophysics Data System Bibliographic Services.


{\it Facility }{\facility Swift (UVOT), Subaru, Spitzer (IRAC) } 

\bibliographystyle{apj}

\bibliography{apj-jour,bibliography_file}

\end{document}